\documentstyle[psfig]{mn2e}
\def\ltsima{$\; \buildrel < \over \sim \;$}
\def\simlt{\lower.5ex\hbox{\ltsima}}   
\def\gtsima{$\; \buildrel > \over \sim \;$}
\def\simgt{\lower.5ex\hbox{\gtsima}}

\input{epsf}

\title{The formation and evolution of bars in low surface brightness galaxies
with cold dark matter halos.}

\author{Lucio Mayer $^1$ \& James Wadsley $^2$\\$^1$ Institute for Theoretical
Physics, University of Zurich, Wintherurestrasse 190, 8057 Zurich, Switzerland\\$^2$Department of Physics \& Astronomy, McMaster University, 1280 Main St. West
, Hamilton ON L8S 4M1 Canada}

\date{\today}

\begin{document}

\maketitle

\begin{abstract}

We perform several high resolution N-Body/SPH simulations of low surface
brightness galaxies (LSBs) embedded in cold dark matter halos to study how 
likely is  bar formation in such systems.
The behavior of various collisionless galaxy models is studied both 
in isolation and in the presence of a large perturbing satellite. 
We also consider models with a dominant gaseous component in the disk.
We find that in general bar formation requires disk masses at least a 
factor of 2 higher
than those inferred for LSBs under the assumption of a normal
stellar mass-to-light ratio. Low surface density stellar disks contributing
less than 10\% of the total virial mass are 
stable within NFW halos spanning a range of concentrations.
However, a purely gaseous disk can form a bar even for quite low 
masses and for realistic temperatures provided that cooling is very 
efficient (we adopt an isothermal equation of state)
and that the halo has a very low concentration, $c < 5$.
The bars that form in these LSB  models are shorter than the 
typical halo scale radius - their overall angular momentum content
might be too low to affect significantly the inner dark halo structure.  
Once formed, all the bars evolve into bulge-like 
structures in a few Gyr and can excite spiral patterns 
in the surrounding disk component. 
The recently discovered red LSBs show significant non-axisymmetric structure 
and bulge-like components and share many of their structural properties
with the final states of our LSB models with massive disks. 
Our results imply that a bulge-like component must be present in any low
surface brightness galaxy that ever went bar unstable in the past.

\end{abstract}

\begin{keywords}{galaxies: dynamics --- galaxies: evolution --- galaxies: 
halos ---methods: N-Body simulations ---cosmology:theory}

\end{keywords}

\section{Introduction}

Deep surveys carried out in the last decade or so have unveiled a vast population of low surface brightness disk galaxies (LSBs) that range from dwarf  to fairly luminous galaxies and even giant spirals (Bothun, Impey \& Malin 1986;Bothun et al. 1987; Schombert et al.1992; Impey et al. 1996). Typical 
LSBs are late-type objects characterized by blue colors, low metallicity, 
very high gas fractions and low levels of star formation activity.
More recently, a new class of gas-rich, late-type 
red LSB galaxies has been discovered (O'Neil et al. 2000) together 
with another population of red, early-type LSBs (Bejersbergen, de Blok \& Van 
der Hulst 1999; Galaz et al. 2002). Some of the red LSBs have 
metallicities close to the solar value (Bergmann, Jorgensen \& Hill 2002).
Hence LSBs probably span a range of properties comparable to  
"normal" high surface brightness spiral galaxies (HSBs).  
The high gas fractions of LSBs suggest that they are quite unevolved objects with very low past and present star formation rates. The low metallicity (de Blok \& Van der Hulst 1998a,b) and the low gas surface densities (de Blok, 
McGaugh \& Van der Hulst 1996a) can explain the low star formation rates 
of blue LSBs (Gerritsen \& de Blok 1997; Van den 
Hoek et al. 2000). The fact that LSBs and HSBs seem to follow the same Tully-Fisher relation (Zwaan et al. 1995; McGaugh et al. 2000) implies that the former have much higher dark matter
contents than the latter within the region typically probed by the
observations (of order of a few disk scale lengths). This conclusion
stems from the low stellar mass-to-light ratios suggested by the blue 
colors of most LSBs. The latter imply that the low surface brightness
is really the product of a low stellar surface density (de Blok, Van
der Hulst \& McGaugh 1996b) -
a higher dark matter density is then required to explain why these 
galaxies have circular velocities comparable to those of similarly 
luminous HSBs (Verehijen \& de Blok 1999). 
Hence LSB galaxies offer an ideal opportunity to study the distribution of dark matter and compare the observational results with current theories of structure formation. 

Predictions of cold dark matter simulations were found to disagree with the rotation curves measured from HI emission
in many dwarf galaxies and LSB galaxies (Moore 1994; Flores \& Primack 1994)). 
Observed galaxies would have halos with a constant density core whereas simulated halos have cuspy density profiles falling as $r^{-1}$  (the 
NFW model, Navarro, Frenk \& White 1997) or even steeper (Moore et al. 1999a).
While some observers have pointed out several limitations of the original measurements
of rotation curves based on HI emission from atomic hydrogen and have thus analyzed several of the 
same galaxies using stellar H$\alpha$ line (Swaters et 
al. 2000; Van den Bosch \& Swaters 2001; Swaters 2003), recent high resolution data obtained  using both techniques seem to confirm that the majority of the rotation curves of LSB
galaxies cannot be fit by the cuspy halo profiles  predicted by CDM cosmogonies
(de Blok, McGaugh \& Rubin 2001a,b;de Blok et al. 2001;de Blok \& Bosma 2002). 
In dwarf galaxies strong supernovae feedback might play a role in affecting the overall
mass distribution (Navarro et al. 1994), but the quiescent star formation 
histories of LSB galaxies together with the fairly large potential wells of many of them rule out 
such a scenario (Bell et al. 1994, 1999; Bergman, Jorgensen \& Hill 2002).

Recently Weinberg \& Katz (2002) have suggested that the dark matter cusp might
be erased thanks to the dynamical interaction with a stellar bar.
It has been known for a while that a rotating barred potential would slow down due to dynamical friction against the halo
background, shedding its angular momentum to the latter (Hernquist \& Weinberg 1994). 
Of course, the deceleration of the bar will be stronger in more massive halos.
Debattista \& Sellwood (1998, 2000) where the first to note that the dynamical interaction 
between the bar and the halo could provide clues to the nature of the latter; they noted that
if galaxies have massive dark halos as predicted by cold dark matter models it
would be
hard to explain why bars in many galaxies are quite fast rotators.
On the other end, Valenzuela \& Klypin (2002) claim that the net transfer
of angular momentum to the bars is quite weak - bars slow down on a quite
long timescale even in CDM halos as angular momentum is exchanged back 
from the halo to the individual stellar orbits supporting them once a 
sufficiently high force resolution is used in the simulations.
Weinberg \& Katz (2002) consider the idealized case 
of a very massive non-responsive bar and find that the resonant transfer of energy  and
angular momentum from the bar to an NFW halo can actually change the density 
profile of the latter creating a constant density in only a few Gyr. 
The effect would occur only when the resonant orbits in the halo are well
resolved, which requires a very large number of particles in an N-Body 
simulation, of order of a few millions.
But Sellwood (2002), using a different numerical technique, finds
the effect to be almost negligible for more realistic initial
setups of the galaxies in which the bar is not imposed from the start and
can evolve instead of being just a rigid potential.
Whereas further
investigation of the effectiveness of the bar-halo interaction with self-consistent stellar bars is necessary, an even more basic question arises;
can this mechanism be at play in the most interesting case, namely
that of LSB galaxies? How likely is that these galaxies would
go bar unstable? Indeed, numerical studies on bar formation and bar-halo interactions have always employed models of high surface brightness galaxies.

While bars seem to be ubiquitous among spiral galaxies as a whole (Eskridge et al. 2000),
low surface brightness galaxies are expected to be  stable to bar formation due to a combination of
low disk self-gravity and high dark matter contents 
(Mihos et al. 1997). But the current status of observations seems to suggest a quite complex scenario.
In fact, while blue LSB disk galaxies are indeed typically non-barred
(although some dwarf LSBs have Magellanic bars),  red LSB galaxies comprise several 
systems showing distortions, evident bars and even bulge-like components (Bejersbergen et al. 1999).
Tidal encounters might also trigger bar formation or other non-axisymmetric distortions even
in LSBs (Mihos et al. 1997) and might actually be required to destabilize the 
gas in the disks and sustain star formation (Schombert et 
al. 2001; Verde et al. 2002).

In this paper we study the formation of bars in models of LSB galaxies
comprising a stellar and/or gaseous disk embedded in a dark matter halo
whose structure is motivated by the results of CDM models. 
Such an approach
is novel among studies of bar formation in galaxies, having been only 
partially adopted by Valenzuela \& Klypin (2002).
We have carried out a large set of high resolution N-Body/SPH 
simulations using the parallel binary tree + SPH code PKDGRAV/GASOLINE (Stadel, 2001;
Stadel, Richardson \& Wadsley 2002; Wadsley, Stadel \& Quinn, 2003).

Weinberg \& Katz (2001) argue that, although present-day low surface brightness
galaxies might be stable to bar formation, they could have undergone a bar instability 
in the past when they first assembled a cold, massive disk. Although
such a bar would have likely triggered a burst of
star formation, this being difficult to reconcile with the many hints pointing
towards a fairly smooth star formation history for these galaxies (Bergman,
Jorgensen \& Hill 2002), observations still cannot exclude that at least the redder
among these galaxies might hide a faded massive disk (O'Neil et al. (2000)).
The bulge-like components seen in many red LSBs might be the
result of secular evolution of an old bar (Combes 1991; Carollo et al 2001).
In brief, the origin and nature of LSB galaxies is still subject to debate
and therefore we will consider a range 
of models  covering a vast parameter space in terms
of masses and internal structure of disks and halos. Our only two 
constraints will be that the 
rotation curves of the models have to resemble those of observed LSBs, i.e. they have to be slowly rising 
out to several disk scale lengths (de Blok \& McGaugh 1997), and that disks have scale lengths 
bigger than those of HSBs having the same luminosity (Zwaan et al. 1995).

Models with a major gaseous disk can represent either 
LSBs during
their early evolutionary stage or those numerous present-day LSB galaxies in
which the baryonic mass is mostly contributed by the gas component.
(de Blok \& McGaugh 1997).
We note that the same disk model realized as purely gaseous instead of purely stellar
is not expected to have the same stability properties; it has been shown that fluid 
configurations tend to be more stable because pressure is generally isotropic while the corresponding stellar analog, 
the velocity dispersion, is generally anisotropic (Cazes \& Tohline 2000). 
However, the study of the stability of
purely gaseous disks embedded in dark matter halos is new to our knowledge; works
exist on the stability of generic axisymmetric or triaxial fluid configurations 
(Cazes \& Tohline 2000, Barnes \& Tohline 2001)
and even on bar-unstable exponential gaseous galactic disks (Friedli \& Benz (1993; 1995) 
but all these studies did not include any realistic dark matter component 
in their models
and are therefore extremely idealized when applied to the dynamical evolution 
of real galaxies.
The stability of gaseous and stellar systems, both uniformly rotating and
differentially rotating, was also studied by Christodoulou et al. (1995) -
these authors tried to extend to gaseous disks within dark halos a
criterion for stability
against bar formation originally developed by Efstathiou, Lake \& Negroponte (1982) for stellar disks but 
did not test their conclusions with numerical simulations.

Finally, we will also consider the case of perturbations tidally induced by massive satellites, as expected during the hierarchical build-up of structures.
Strong tidal interactions with even more massive galaxies would also occur in a
hierarchical scenario and surely will drive bar formation but they would also 
transmute  these fragile disk galaxies into spheroidals or even destroy them 
(Moore et al. 1999; Mayer et al. 2001a,b).

The paper
is organized as follows; in section 2 we describe the models and the initial
setup of the simulations, in section 3 we illustrate the results of the
simulations, section 4 contains the discussion and a summary follows in the last 
section.

\section{Galaxy models and simulations}

Galaxy models are built as in Mayer et al. (2001b;2002) using the technique originally developed
by Hernquist (1993) (see also Springel \& White 1999).
We use a system of units such that $G=1$, $[M]=6.5 \times 10^9
M_{\odot}$ and $[R]=6$ kpc.
The models comprise a dark matter halo and an embedded stellar or gaseous 
disk (or both). Structural parameters were chosen in order to obtain slowly rising  rotation curves resembling those
published for LSB galaxies (e.g. de Blok et al. 
2001a,b, de Blok, McGaugh \& Rubin 2001, de Blok \& Bosma 2002, 
Van den Bosch \& Swaters 2001). We start by choosing the value of the circular velocity of
the halo at the virial radius, $V_{vir}$, which, for an assumed cosmology 
(hereafter $\Omega_0=0.3$, $\Lambda=0.7$, $H_0=65$ kms$^{-1}$Mpc$^{-1}$) automatically
determines the virial mass, $M_{vir}$, and virial radius, $R_{vir}$, of the halo
(Mo, Mao \& White 1998). We choose $V_{vir} = 75$ km/s.
Halos have NFW density profiles (Navarro, Frenk \& White 1995, 1997) with  different halo 
concentrations,$c$, and spin parameters, $\lambda$ (Mo,Mao \& White 1998).
The 
concentration is defined as $c=R_{vir}/r_s$, where $r_s$ is the halo 
scale radius; the spin parameter is defined as $\lambda= J{|E|}^{1/2}G^{-1}{M_{vir}^{-5/2}}$, 
where $J$ and $E$ are, respectively, the total angular momentum and total energy of the halo and
$G$ is the gravitational constant.

The value of the concentration 
$c$ basically defines what fraction of the total
mass of the halo is contained within its inner regions, where the baryonic
disk lies; the concentration increases with decreasing mass 
and, for a given mass, has a scatter of roughly a factor of 2, mainly due to different formation
epochs (Bullock et al. 2001; Eke, Navarro \& Steinmetz 2001). 
The average concentration for galaxies with $V_{vir} \sim$ 75 km/s is
$\sim 12$ in the standard LCDM model assumed here.
However fitting rotation curves of LSBs with NFW profiles often requires 
$c \simlt 5$ (Van den Bosch \& Swaters 2001; de Blok \& Bosma  2002),
at  the lower end of the allowed range of values.
We consider three values for the halo concentration, $c=4$, $c=7$ and $c=12$.

Placing galaxy disks inside cosmological halos would in principle require to
have first solved the problem of galaxy formation in the LCDM scenario.
Incidentally, whereas for years simulations of galaxy formation within CDM
models have been plagued by the so-called angular momentum "catastrophe" (Navarro
\& Steinmetz 2000), producing
only tiny disks an order of magnitude smaller than those of real galaxies, new 
SPH simulations with considerably higher resolution and an improved treatment of
gasdynamics find this problem to be significantly alleviated
(Governato et al. 2002, see also Sommer-Larsen et al. 2002; 
Thacker \& Couchman 2001). These new simulations
produce disks bearing scaling relations with their dark matter halos reasonably
close to the predictions of the semi-analytical models by Mo, Mao
\& White (1998) - the latter are able to match the properties of spiral galaxies
assuming angular momentum conservation of the baryons as they cool into
the halos and settle into a centrifugally supported disk(Fall \& Efstathiou 1980).
We thus construct models of LSB galaxies that follow the scaling
relations of Mo, Mao \& White (1998). The procedure used to assign structural
parameters is described in detail in Mayer et al. (2001b) and Mayer et al. (2002). Here we recall that we use 
exponential disks (with a few exceptions indicated below) and that their mass and scale length are determined primarily by $M_{vir}$ and $\lambda$ and, to a minor extent, by
the disk/halo mass ratio, $f_d$, and halo concentration $c$ (the latter two
parameters contribute to specify the potential energy of the system and thus
the rotational energy needed for centrifugal support). The adiabatic contraction of the halo 
in response to the accumulation of baryons at the center is also taken into account
(Springel \& White 1999).
As we mentioned in section 1, rotation curves of LSB
disk galaxies suggest that these systems are extremely dark matter dominated, with
stellar disk/halo ratios $\sim 0.03$ or less, namely at least a factor of 2 lower than those 
of HSBs of comparable luminosity (O'Neil et al. 2001, Chung et al. 2002). 
However, this difference probably reflects the fact that in LSBs a larger fraction of the
baryons, sometimes most of them, are found in the gaseous component; indeed the total disk
mass (stars + gas) relative to the halo mass is quite similar in LSBs and HSBs (McGaugh et
al. 2000). Previous works on the detailed mass modeling of LSB galaxies have indeed found 
a typical value of $0.065$ for the total disk/halo mass ratio in these systems (Hernandez
\& Gilmore 1998).
We thus adopt the view that LSBs are simply more extended than HSBs because
of larger halo spin parameters and eventually lower halo concentration, as 
originally suggested by Dalcanton et al. (1997) and Mo, Mao \& White (1998), 
(see also Jimenez et al. (1997) and Hernandez \& Gilmore (1998)) and as 
confirmed more recently by Jimenez, Verde \& Oh (2003) in their modeling of
H$\alpha$ rotation curves.
The disk mass fraction $f_d$ is usually set equal to $0.05$ or less in our models but we also consider
cases in which $f_d=0.1$, such a massive disk being expected  
in scenarios  where LSBs have a massive faded stellar component
(O'Neil et al. 2001;Chung et al. 2002; Galaz et al. 2002). The highest value of $f_d$ is still 
consistent
with  the upper limit set for $\Omega_b$ 
by nucleosynthesis, $\sim 0.13$ (Fukugita, Hogan \& Peebles 1999); in this case, however, we
are implicitly assuming that nearly all the baryons ended up in the disk
(it is somewhat an extreme assumption - see Verde et al. 2002).
Jimenez et al. (1998), who coupled a stellar population synthesis
model to a simple galaxy formation scheme in the context of the hierarchical
clustering, find that $\lambda > 0.05$ is required to match the
surface brightness and colors of several "blue" LSBs. 
In our models, halos have spin parameters either $0.065$ or $0.1$, larger than the
mean value found in cosmological simulations, $\lambda \sim 0.04$ 
(e.g. Gardner 2001, Lemson \& Kaufmann 1998) - these values yield a
disk scale length, $R_h$, in the range 2-6 kpc, consistent with observations (e.g. Zwaan et al. 1995,
O'Neil et al. 1998).
Although our halos have $V_{vir} \sim 75$ km/s, the different halo concentrations 
and the addition of the stellar disk 
produce a peak rotational velocity, $V_{peak}$ (at $\sim 2 R_h$)
between $95$ and $130$ km/s, within the range of the maximum rotation speeds measured
for the majority  of LSBs in the samples 
by de Blok \& McGaugh (1997) and by de Blok \& Bosma (2002).
Indeed it is only $V_{peak}$ which is accessible to observations based
on HI or H$\alpha$ kinematics, not $V_{vir}$, which is too far out in 
the halo. 
The rotation curves of some galaxy models are shown in Figure 1.

The setup of the stellar disk is complete once even the Toomre parameter,
$Q(R)$, is defined (Toomre, 1964)
This corresponds to fixing the local stellar 
velocity dispersion $\sigma_R$, as $Q(R)=\sigma_R \kappa/3.36G \Sigma_s$,
where $\kappa$ is the local epicyclic frequency , $G$ is the gravitational
constant and $\Sigma_s$ is the disk surface density.  
The $Q$ parameter at $\sim 1$ disk scale length will 
be given as a reference; the latter is 
typically set equal to 1.2 (see Table 1 for the precise values adopted). $Q$, 
in combination with other parameters (Binney \& Tremaine 1987; Mihos et al. 1997), 
determines the stability of the disks to bar formation, Numerical studies done in the 
past suggest that $Q \simgt 2$ is needed for stability against 
stellar bar growth in isolated galaxies (Athanassoula \& Sellwood 1986; 
Friedli 2000); however these studies were conducted using galaxy models 
of "normal" HSB spiral galaxies, while 
our LSB models have a lower disk self-gravity and could well be stable even 
for smaller $Q$ values (Mihos et al. 1997).

\begin{figure}
\epsfxsize=10truecm
\epsfbox{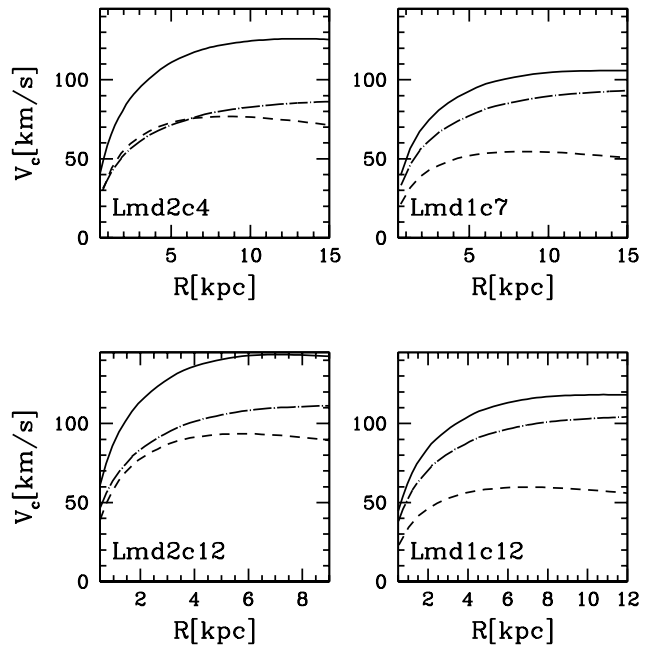}
\caption{Rotation curves of galaxy models with purely stellar disks out to 
three disk scale lengths.  The solid line denotes
the total curve, while the dot-dashed and the dashed lines represent the 
separate contribution of, respectively, dark matter and stars. The name of 
each model is indicated in the panels (see Table 1).}
\end{figure}

We note that models with $c=4-7$ , $\lambda=0.065$ and $f_d 
=0.05$ have a B band central surface brightnesses, ${\mu_0}_B$, comprised between $22.5$ and $23.5$ mag arcsec$^2$ 
for a B band stellar mass-to-light ratio ${(M/L_B)}_* =2$. 
We avoid more extreme values of the surface density/brightness on purpose; indeed out first
objective is to investigate if the bar instability is possible at least among LSB galaxies
at the bright end of the surface brightness distribution (de Blok \& McGaugh 1997).
The models with massive disks ($f_d=0.1$) would have a 
central surface brightness 0.5 mag lower than the threshold often used  as a 
definition of  LSB galaxies, ${\mu_0}_B  = 
22.5$ mag arcsec$^{-2}$, for a stellar mass-to-light ratio $\sim 2$,
but, if representative of faded disks with $M/L_* \simgt 5$ (O'Neil
et al. 2001), would have  $\mu_b \ge 23.5$ mag arcsec$^{-2}$.

\begin{table*}
\centering
\caption{Parameters of the initial models.
Column 1: Name of the model.
Column 2: Disk/halo mass ratio. Column 3: 
halo concentration. Column 4: halo spin parameter.
Column 5: disk gas fraction. Column 7: type
of gas surface density profile. Column 8: $\varepsilon_d$ stability
parameter. Column 9: Toomre Q parameter at 1 disk
scale length; Column 10: $X_2$ parameter
for swing amplification at 1 disk scale length. Column 11:
mass of the satellite relative to total virial mass of the system.
Column 12: whether the system forms or not a bar ("tr" stands for transient)}
\begin{tabular}{lcccccccccc}
Model &  $f_d$ & $c$ & $\lambda$ & $f_g$ & gas profile & $\varepsilon_d$ & Q & $X_2$ & $M_{sat}/M_{vir}$ &bar  \\
\\
Lmd0c4  & 0.038 & 4 & 0.065 & 0 & 0 & 1.123 & 1.2 & 2.6 & 0 & no\\
Lmd1c4	 & 0.05 & 4 & 0.065 & 0 & 0 & 0.945 & 1.2 & 1.9 & 0 & no\\
Lmd1c4Q2  & 0.05 & 4 & 0.065 & 0 & 0 & 0.945 & 1 & 1.9 & 0 & no\\
Lmd1c4Q3   & 0.05 & 4 & 0.065 & 0 & 0 & 0.945 & 0.5 & 2 & 0 & tr\\
Lmd1c7      & 0.05 & 7.5 & 0.065 & 0 & 0 & 1 & 1.2 & 2.42 & 0 & no\\
Lmd1c12        & 0.05 & 12 & 0.065 & 0 & 0 & 1.02 & 1.2 & 2.47 & 0 & no \\
Lmd2c4	    & 0.1 & 4 & 0.065 & 0 & 0 & 0.703 & 1.2 & 1.36 & 0 & yes \\
Lmd2c12  	& 0.1 & 12 & 0.065 & 0 & 0 & 0.704 & 1.2 & 1.55 & 0 & yes\\
Lmd2c12b	& 0.1 & 12 & 0.1 & 0 & 0 & 0.909 & 1.2 & 2.3 & 0 & yes \\
Lmd2c12c      & 0.1 & 12 & 0.065 & 0 & 0 & 0.704 & 0.5 & 1.55 & 0 & yes \\
Lmd1c4g	     & 0.05 & 4 & 0.065 & 1 & exp & 1 & 1.2 & 1.9 & 0 & yes \\
Lmd1c4gb     & 0.05 & 4 & 0.065 & 1 & exp & 1 & 1.5 & 1.9 & 0 & no \\
Lmd2c4g      & 0.05 & 4 & 0.065 & 1 & exp & 0.703 & 1.2 & 1.36 & 0 & yes \\
Lmd1c12g	& 0.05 & 12 & 0.065 & 1 & exp & 1.02 & 1.2 & 2.47 & 0 & no \\
Lmd2c12g	& 0.1 & 12 & 0.065 & 1 & exp & 0.707 & 1.2 & 1.55 & 0 & yes \\
Lmd1c4sg	& 0.05 & 4 & 0.065 & 0.5 & exp & 1.41 & 2.4 & 3.8 & 0 & no \\
Lmd1c4gc	& 0.05 & 4 & 0.065 & 1 & const. & 1.4 & 2.2 & 10 & 0 & no\\
Lmd1c12gc	& 0.05 & 12 & 0.065 & 1 & const. & 1.2 & 1.5 & 13 & 0 & no \\
Lmd1c4sat	& 0.05 & 4 & 0.065 & 0 & 0 & 0.945 & 1.2 & 2 & 0.03 & no\\
Lmd1c4Q3sat     & 0.05 & 4 & 0.065 & 0 & 0 & 0.945 & 0.5 & 2 & 0.03 & tr\\
Lmd1c12sat	& 0.05 & 12 & 0.065 & 0 & 0 & 1.02 & 1.2 & 2.47 & 0.03 & no\\

\end{tabular}
\label{t:simul}
\end{table*}

\begin{figure}
\epsfxsize=10truecm
\epsfbox{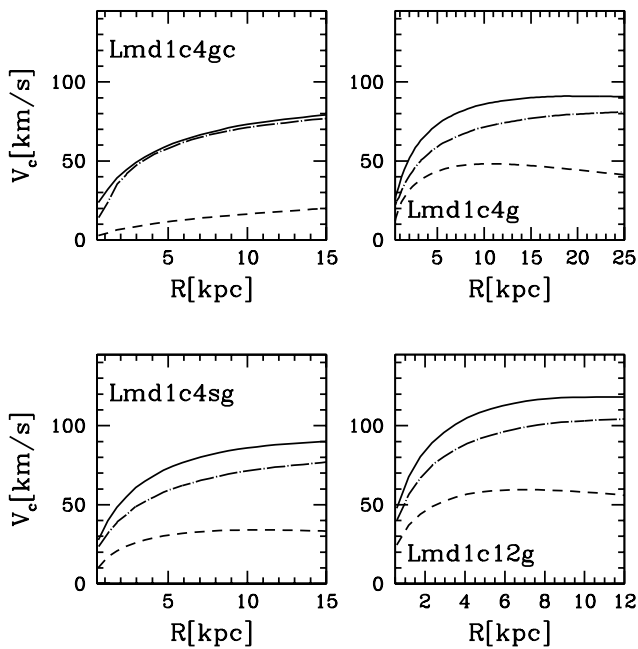}
\caption{Rotation curves of galaxy models with purely gaseous disks out to three
disk scale lengths.
The solid line denotes
the total curve, while the dot-dashed and the dashed line represent the separate
contribution of, respectively, dark matter and gas. The name of each model is
indicated in the panels (see Table 1).}
\end{figure}

In the runs employing a gaseous disk this has
a temperature of  7500 K; the kinematics of the neutral hydrogen in the disks of spiral galaxies 
yield typical velocity dispersions consistent with this
temperature (Martin \& Kennicutt 2001).
The gaseous disk has 
either a constant or an exponential surface density profile with the same scale length of the disk
(see Mayer et al. 2001b); in fact some LSB galaxies have rotation curves indicating that the HI component has 
a profile flatter than the stellar disk (e.g. de Blok \& McGaugh. 1997).
In a gaseous disk the Toomre parameter is defined as $Q(R)= v_s \kappa/\pi G \Sigma_g$, where $v_s$ is the sound speed and $\Sigma_g$ is the surface 
density of the gas.
For the assumed disk and halo masses a temperature of 7500 K implies $Q \simgt 1$, namely 
comparable to that of the collisionless runs.

\begin{center}
\begin{figure}
\epsfxsize=5truecm
\epsfbox{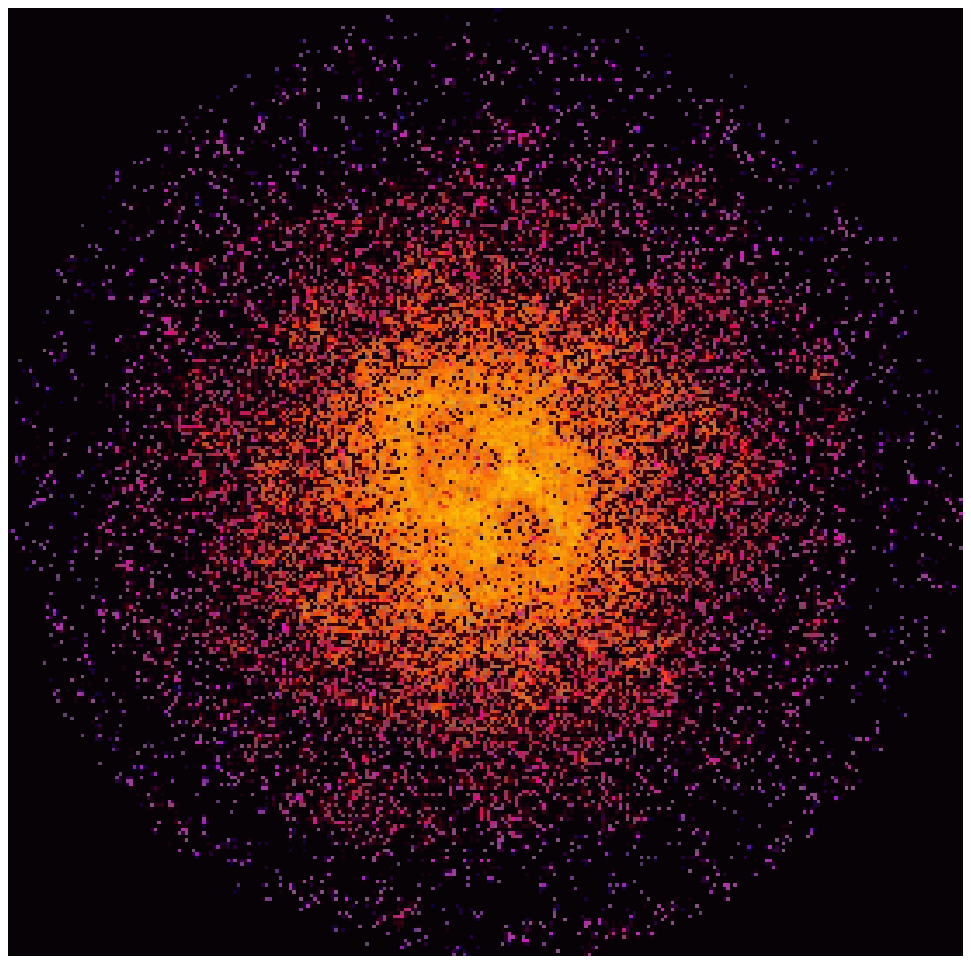}
\epsfxsize=5truecm
\epsfbox{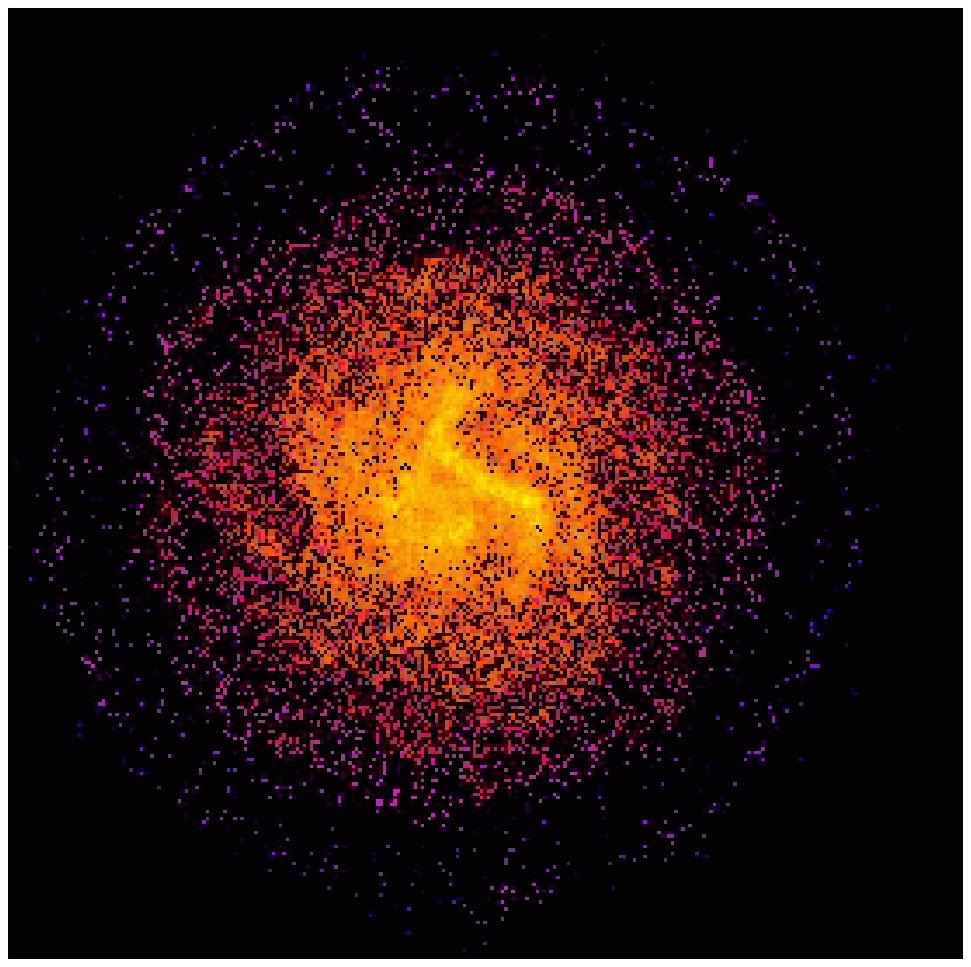}
\epsfxsize=5truecm
\epsfbox{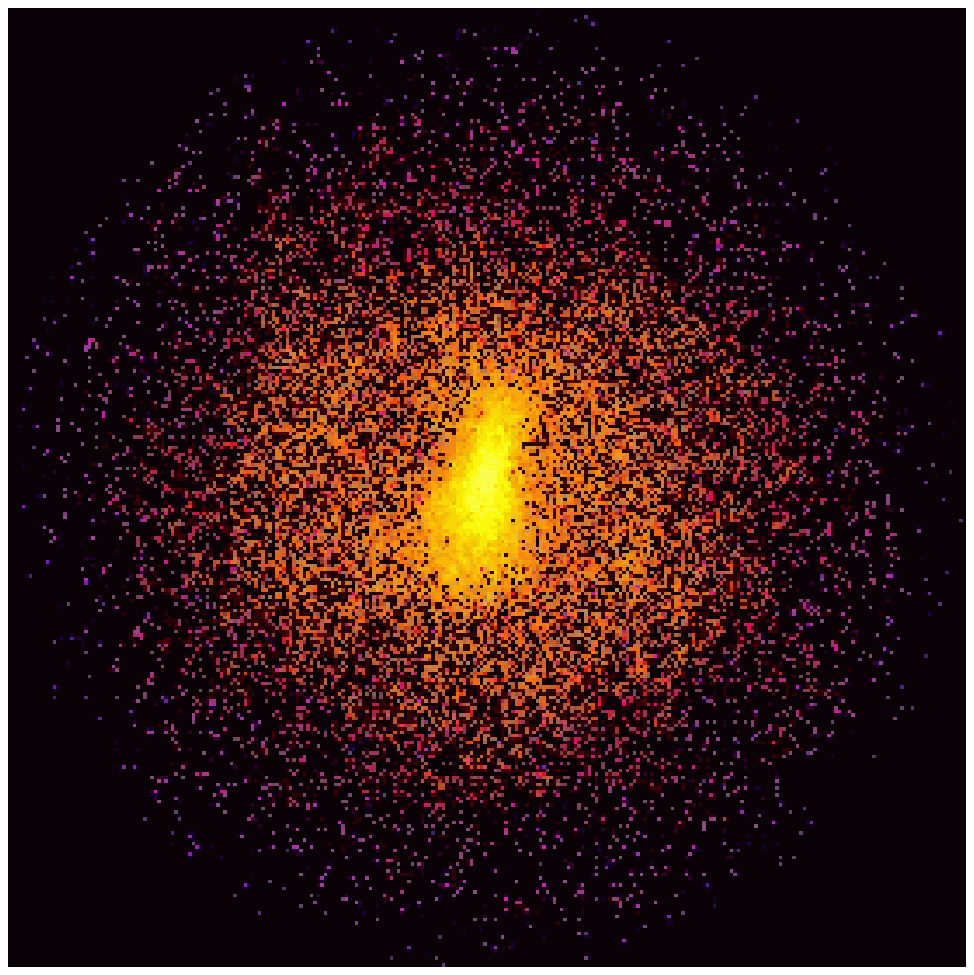}
\epsfxsize=5truecm
\epsfbox{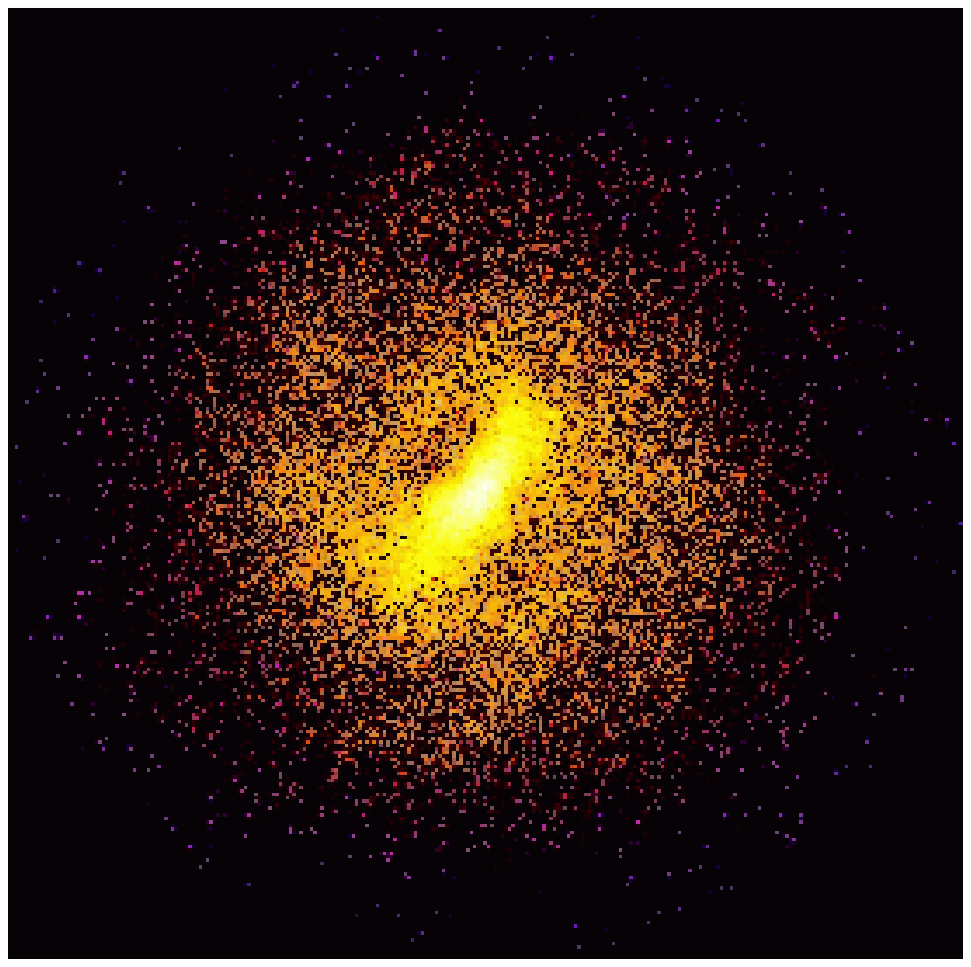}
\caption{Color-coded face-on view of the stellar density after 3 Gyr of evolution for some of 
the collisionless galaxy models (see Table 1). From top to bottom; Lmd1c4, Lmd1c4Q3, Lmd2c4 and Lmd2c12 are shown.
Brighter colors correspond to higher densities; a central bar is evident in some cases. Boxes are 20 kpc on a side }
\end{figure}
\end{center}

We do not include radiative cooling in the treatment of gasdynamics but we adopt
an isothermal equation of state to model dissipation. This is somewhat idealized
as it stands on the assumption that the thermal energy generated by, e.g. shocks and supernovae explosions, 
can be instantly radiated away, i.e. cooling is assumed to be very efficient.
however simulations of galaxy formation and evolution which 
explicitly include both radiative cooling and heating show that the temperature
of the gas in galactic disks stays always close to $10^4$ K (Gerritsen \& Icke 1997; Katz, Hernquist \& Weinberg 1992; Sommer-Larsen et al. 2002; Governato et al. 2002; Hernquist \& Katz 1989; Katz \& Gunn 1991; Navarro 
\& Steinmetz 2000; Governato et al. 2002 - note however that lower temperatures are not allowed 
in many of these works because they usually adopt cooling functions with a 
cut off at about $10^4$ K). 
Supernovae might heat the gas very efficiently
if the energy of their explosions is partly converted into turbulent motions 
(Thacker \& Couchman 2001; Springel \& Hernquist
2002) instead of being entirely converted into thermal energy, but their 
global  impact on galaxies bigger than dwarfs is not yet established both 
observationally and theoretically (Martin 1999; MacLaw \& Ferrara 1999; 
Benson et al. 2003). 
We recall that the general expectation is that LSB galaxies should
be quite stable to non-axisymmetric instabilities; in this context the 
assumption of an isothermal equation of state will provide the most
favorable condition for the formation and survival of bars in gaseous disks 
by forcing the disk to remain cold (for instance it might  underestimate 
heating  when a disk has already entered a 
phase of strong instability) - in reality the same systems can only be more
stable if they retain some of the heat generated during their dynamical
evolution.

However, in order to understand how sensitive are the results to the assumed 
equation of state, we evolve the same initial conditions with both
an isothermal and and
adiabatic equation of state (not shown in Table 1); the adiabatic runs 
represent the situation at the opposite extreme of the isothermal runs,
namely radiative cooling is completely switched off.
A final type of models comprise systems in which the stellar and gaseous component make an equal contribution to the disk mass - this can
reflect an evolutionary stage intermediate between those of models with
purely gaseous and purely stellar disks.
The rotation curves of some of the models with a gas component are shown in Figure 2.

Gasdynamical simulations were carried out with GASOLINE, an extension
of PKDGRAV (Stadel 2001) which uses SPH to solve the hydrodynamical equations (Wadsley, Stadel \& Quinn, 2003; Stadel,
Richardson \& Wadsley 2002). The gas is treated as an ideal gas with equation
of state $P=(\gamma -1)\rho u$, where $P$ is the pressure, $\rho$ is
the density, $u$ is the (specific) thermal energy, and  $\gamma$ is the ratio 
of the specific heats (adiabatic index)
and it is set equal to $5/3$ (we are assuming that the gaseous disk represents
the atomic hydrogen component of the galaxy). In its general form the code solves an 
internal energy equation which includes an artificial viscosity term
to model irreversible heating from shocks.
The code adopts the standard Monaghan artificial viscosity as well as the
Balsara criterion to reduce unwanted shear viscosity (Balsara 1995).
In the isothermal case the thermal energy is constant over time and
so no thermal energy equation is required
(any heating resulting from the artificial viscosity term, which
is still present in the momentum equation, is radiated away),
In the adiabatic case the thermal energy can rise as a result of 
compressional and shock heating and can drop following decompressions.

We also explore the case of LSB galaxies 
tidally disturbed by a massive companion having a mass comparable to
that of the LMC. Other regimes of external tidal perturbations should not be
relevant here.
In fact more massive satellites would rapidly merge at
the center before being substantially stripped (Colpi, Mayer \& Governato 1999;
Taffoni et al. 2003)
and would likely destroy the fragile disks, while repeated fly-byes with
even more massive galaxies in dense environments, like galaxy clusters
or groups, would actually transform LSB galaxies into spheroidals or S0 
systems (Moore et al. 1999; Mayer et al. 2001a,b). For this set of runs we use
LSB galaxies with purely stellar disks and the
satellite is a deformable spherical halo with a fairly concentrated NFW
profile (c=15), a high concentration making the satellite stiffer and 
the associated perturbation stronger. The models used for these runs are
indicated in Table 1.
The satellite moves
in the same direction of the disk rotation (prograde), which can maximize
the tidal effects due to resonances between its orbital frequency
and the frequencies of disk stars (Velazquez \& White 1997). We consider both a very
eccentric (apo/peri=15) and a nearly circular orbit (apo/peri=2), both
having a pericenter of 30 kpc, namely grazing the disk; indeed, while 
repeated tidal
shocks can be effective in driving global instabilities (Mayer et al. 2001b),
near-resonant interactions between the satellite orbital frequency and the
internal frequency of the galaxy might also excite non-axisymmetric modes in the disk
(Weinberg 2000; Weinberg \& Katz 2002).

For the models with a single-component disk we use 500,000 or 1 million particles
for the halo  and 50,000 particles for the
disk (either gaseous or stellar). In the models with a two-components disk
we use 25,000 particles for both components and the same number of halo particles
as before.
In order to test how our results
depend on resolution, we also performed runs in which the latter is doubled
or tripled in either the dark or the baryonic component.
The resolution used for the halo is more than adequate to study physical
bar formation - previous work showed that, with a few hundred 
thousand particles in the halo, spurious bar modes cannot be triggered in 
otherwise stable galaxies (Mayer et al. 2001b). 
The only other existing work in which N-body simulations of bar formation 
with similarly high resolution were performed is Valenzuela \& Klypin (2002);
compared to them we use a lower force resolution, i.e we employ
a (spline) softening corresponding to $0.05 R_h$ (for both the stars and 
the gas) while  they use $0.01R_h$. However, a small softening 
is not 
necessarily a good choice - force resolution must be balanced with mass
resolution in order to avoid increasing the noise due to the discrete
representation of the systems (Hernquist, Hut \& Makino 1993;Moore,
Katz \& Lake 1996) and our conservative choice has been widely tested in 
this respect (Mayer et al. 2001b).
Yet, in order to test how are results depend on force resolution, we have 
also run selected models a second time using the same softening as in 
Valenzuela \& Klypin (2002) without finding any significant differences in the results.

\begin{figure}
\epsfxsize=8truecm
\epsfbox{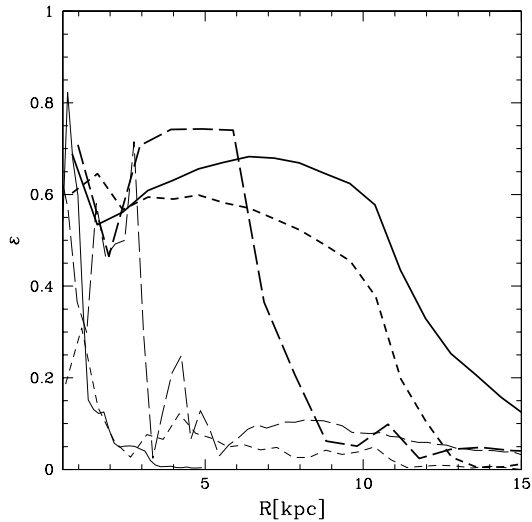}
\caption{Ellipticity parameter of the stellar component in some of the 
collisionless simulations after 3 Gyr of evolution.
The thick solid line refers to model Lmd2c12, the thick dashed line to model
Lmd2c4, the thin dashed line to model Lmd1c7, the thick long-dashed line to model Lmd1c4Q3, the thin solid line to model Lmd1c12 and the thin long-dashed
line to model Lmd1c4.}
\end{figure}

\section{Results}

We evolve the various initial galaxy configurations for up to 10 Gyr, which
corresponds to several disk dynamical times (this being of the order of a few
times $10^8$ years) and already represents a significant fraction of 
the cosmic time. 
The developing of non-axisymmetric structure in our galaxy models can be
seen in the projected color coded density maps shown in this section. We
also quantify the strength of such distortions by means of the ellipticity
parameter $\epsilon=1 - b/a$, where $a$ and $b$ are the major and intermediate axis of the baryonic component of the galaxies (measured on the basis of the 
principal mass moments, $\epsilon=0$ if the galaxy is axisymmetric).
The ellipticity has been found to be a good measure of the strength of bars
in a large number of studies as it correlates with more physical parameters
like the ratio between the radial and the axisymmetric components of the
force (Laurikainen et al. 2002; Das et al. 2001; Laurikainen, Salo 
\& Rautiainen 2002;Martin 1995;Regan \& Elmegreen 1997). Although both bar 
mass and ellipticity would be needed to fully specify the bar contribution to the overall 
stellar potential of the galaxy, the latter is the only parameter easily accessible by the observations (Martin 1995).

\begin{figure}
\epsfxsize=8truecm
\epsfbox{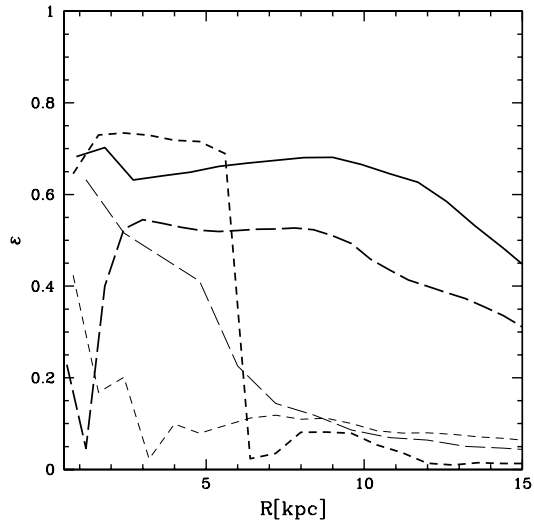}
\caption{Ellipticity parameter of the gaseous component in gasdynamical
runs. Models Lmd2c12g (thick solid line for the isothermal simulation
and thick long-dashed for the adiabatic simulation), Lmd1c4g (thick dashed
line for the isothermal simulation and thin dashed for the adiabatic 
simulation) and Lmd2c4sg (thin long-dashed line) are shown after, respectively,
after, 3, 5 and 3 Gyr (i.e. at the time where the strongest non-axisymmetry is
observed in each of them).} 

\end{figure}

\subsection{Collisionless runs}

LSB models with either low or high concentration and  $f_d < 0.1$
do not undergo any bar instability if $Q > 1$ (taking $Q$ at roughly  1 disk
scale length as a reference like in Table 1). Stability of these models should be even more robust at
higher resolution as numerical noise is suppressed and spurious non-axisymmetric modes induced by the  discrete representation of the collisionless fluid 
gradually disappear (Hernquist 1993). 
A transient bar appears after 3-4 Gyr in LSB models with
 $c=4$ halos and $f_d \simgt 0.05$ only when $Q$  is as low as 0.5 (see Table 1). 
The bar scale length is small, roughly $0.5 R_h$. 
The results for such "light" disk models are shown in Figure 3 and 4.
Instead, models with massive disks, $f_d=0.1$, become bar unstable 
regardless of halo concentration and the value of $Q$.
With such high disk masses bars
are strong and long-lived (bottom
panel of Figure 3, Figure 4 and 10).
Bars form rapidly, after about 2 Gyr, of order of a few disk dynamical
times at $R=R_h$, and are $\sim 1 R_h$ in low concentration halos, and much
longer, $\simlt 2 R_h$, in high concentration halos.
We tested that our results are not compromised by
numerical effects, especially two-body scattering between 
stellar particles and more massive halo particles (Moore et al. 1996) and
lack of force resolution that might bias the evolution of the galaxy
potential (Dehnen 2001); we found good convergence when we
compared a)runs with 500,000 particles halos with runs having 1 million 
particles halos and b)when we compared the latter runs
with runs having the same number of particles but a force resolution 
five times higher (i.e. a softening five times smaller) in both the
dark matter and the stellar component.

At higher values of the halo concentration a higher shear is present 
near the center and swing amplification of $m=2$
modes is expected to be weak unless $Q$ is extremely low 
(Binney \& Tremaine 1987).
The susceptibility of a differentially rotating thin disk to swing 
amplification of some particular mode $m$  can be measured by the 
parameter $X_m={R{\kappa}^2/2\pi m G \Sigma}$ (Toomre, 1981).
 For $m=2$ modes,
$X_2 < 3$ is required for strong swing amplification in 
potentials associated with flat rotation curves (Binney \& Tremaine 1987) but
a lower threshold,  $X_2 < 2$, has been found for systems with slowly
rising rotation curves as those of our LSB models (Mihos et al. 1997; 
Mayer et al. 2001b).
The trend of increasing stability with increasing concentration is
consistent with the fact that the $X_2$ parameter at $R=R_h$ 
is nearly twice as big for halos with $c=12$ compared to halos with $c=4$
with having the same $f_d$ (see Table 1). $X_2$ is indeed
below the expected threshold for instability only for the models with
$f_d=0.1$.
Once initiated, swing amplification cannot proceed if a strong Lindblad inner
resonance is present that cuts off the feedback cycle (Binney \& Tremaine 1987;
Sellwood \& Evans 2001).This can be another way by which galaxies can avoid to develop a bar. 
We checked the height of the Lindblad barrier for different
models. We conclude that stellar frequencies that can be swing 
amplified always lie above the inner Lindblad resonance and hence
this mechanism is negligible in our simulations,
although the 
barrier is certainly higher when a high concentration halo is
present. 
In brief,  we think that our results mostly reflect
the structural difference between systems in which swing amplification can 
start and systems in which it cannot start and that such difference is 
due to a combination of halo concentration and disk mass (both these 
two quantities enter the definition of $X_2$ through, respectively,
$\kappa$ and $\Sigma_d$).
We observe a weak spiral structure even in models with high concentration 
and $f_d=0.05$ and with up to a million particles (as shown by the non-zero
ellipticity of the corresponding models in Figure 4). 
- however  we 
cannot exclude that these features are partially driven by numerical noise
(Hernquist 1993).

We find that the $\varepsilon_d$ parameter
introduced by Efstathiou, Lake \& Negroponte (1983) provides a good
measure of the stability against bar formation. This parameter is defined 
as $\varepsilon_d
=V_{peak}/\sqrt{(GM/R_h)}$, where $V_{peak}$ (see section 2) is 
proportional to the mass of the dark halo within about one
disk scale length, and the 
denominator is proportional to the disk self-gravity.
In Table 1 we list the values of $\varepsilon_d$ for the galaxy models  
and we indicate whether
the galaxy model is found stable or unstable to bar formation.
In Efstathiou, Lake \& Negroponte (1983) it was found that $\varepsilon_d 
\simgt 1.1$ was required for stability against bar formation; we find
that the criterion is somewhat weaker, $\varepsilon_d  \simgt 0.94$.
This slight discrepancy is likely due to the much lower resolution
used in the N-Body simulations of those authors (more than an order of 
magnitude less particles than even the lowest resolution simulations 
presented here) which was likely enhancing bar formation.
The fact that $\varepsilon_d$ does not account for several dynamical
aspects of bar formation, like the stabilizing effect of a large velocity dispersion of the disk and the possible interruption of swing amplification by 
the Lindblad
barriers , and yet it provides a good fit to our results, strengthens
our interpretation that the halo/disk mass ratio within the typical
disk radius, which is related to both halo concentration and
disk mass fraction (see rotation curves in Figures 1 and 2) 
is really determining the behavior of our galaxy models
(had we included models with "hard centers" produced by. e.g., bulges,
the incompleteness of this criterion would have been more manifest,
see Sellwood \& Wilkinson (2001)).
Even so, $\varepsilon_d$ alone provides only a sufficient 
condition for stability to bar formation; a necessary and sufficient condition
(at least for the present models)
can only be obtained by coupling it with another parameter, for instance $Q$, 
and this is shown by comparison between models Lmd1c4, Lmd1c4Q2 and Lmd1c4Q3 
- for the same $\varepsilon_d$ a (transient) instability happens only
when $Q < 1$ (see Table 1 and Figure 3).

\begin{center}
\begin{figure}
\epsfxsize=8truecm
\epsfbox{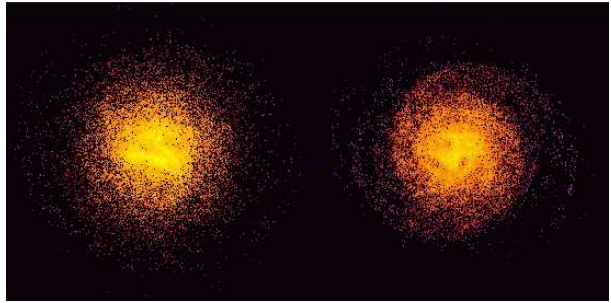}
\caption{Color-coded face-on view of the stellar density of model 
Lmd1c12 (left) and Lmd1c4Q3 (right) perturbed by a satellite on a prograde orbit with, respectively, $apo/peri=2$
and $apo/peri=15$ (see text). Brighter colors represent higher densities. Snapshots are taken after
6 Gyr, when the satellite has performed roughly three orbits. Boxes are 30 kpc on a side.}
\end{figure}
\end{center}

Galaxy models with highly concentrated halos are
stable even when satellites perturb them during several close passages over
10 Gyr (model Lmd1c12sat, see Figure 6).
Models with low concentration halos undergo a bar instability only if the disk has the lowest value of Q (Lmd1c4Q3sat) but this is weak and 
transient exactly as in the case where the galaxy is isolated, and the galaxy only
shows mild non-axisymmetric distortions after several Gyr (Figure 6).
We find these conclusions to hold regardless of whether  
the orbit of the satellite is nearly circular or nearly radial. On a very eccentric orbit the 
impulsive heating
at pericenter drives the formation of outer spiral arms in the disk. 
Therefore, the perturbing satellite does not appear to trigger the  
bar instability in the disks: this is in contrast
to what observed when an LSB satellite interacts with a much more massive galaxy halo (Mayer et al. 2001a, b), suggesting that only extremely 
strong perturbations can stimulate the growth of bar modes 
in these low surface density disks. 

In conclusion there seems to be only a small region
of the parameter space possibly covered by LSB galaxies where 
the formation of a stellar bar is possible within a cuspy halo. In particular,
these disks are stable within NFW halos 
as concentrated as typically expected in LCDM ($c > 10$).
In order to grow bars, models need a baryonic 
fraction as high as  10 \% of the total mass for a spin parameter 
$\sim 0.065$.
This is more than a factor of 2 higher than the baryon fraction estimated
for most observed LSB galaxies ((Hernandez et al. 1998) and is higher by an 
even larger amount compared to the fraction of baryons estimated
to be in stars for a normal stellar mass-to-light ratio 
(de Blok \& McGaugh 1997). Of course
one could imagine that current surveys are still missing a fraction
of the baryons of LSBs 
because these galaxies are very extended and/or
have a significant old stellar population not detectable in the optical
(O'Neil et al. 2000; Galaz et al. 2002); even so the required baryon content 
is quite high, being close to 
$\simgt 0.13$, the upper limit indicated by nucleosynthesis 
(Fukugita, Hogan \& Peebles 1999 - but note that the new WMAP data allow 
for a sensibly higher upper limit, $\sim 0.19$ (Spergel et al. 2001)). 
Hence 
the models with massive disks are somewhat extreme as they demand
that nearly all of the baryons available to the galaxy have already 
cooled into the disk.
As mentioned in section 2, if ${(M/L_B)}_* =2$ model Lmd2c12g has a
central surface brightness high enough to be close to the
upper limit usually adopted as a definition of an LSB galaxy in surveys, 
${\mu_0}_B \sim 22.5$ mag arcsec$^{-2}$, while it would be a more typical LSB 
when placed in a low concentration halo ((${\mu_0}_B \sim 23.5$ mag arcsec$^{-2}$); this is because in our modeling more concentrated 
halos naturally produce more compact disks (Mo, Mao \& White 1998). 
One option to bring down the initial
surface density/brightness (keeping $V_{vir}$ and $f_d$ fixed) is to
start from a larger value of the spin parameter, 
which would reduce the disk surface density by spreading the
same disk mass over a larger radius.
To explore the latter hypothesis we evolved
a galaxy with $\lambda=0.1$ and $c=12$ (model Lmd2c12b, see Table 1)
that would have a lower central surface 
brightness ($\simgt 23.5$ mag arcsec$^{2}$), near the average 
for LSBs. Such  a model still goes bar 
unstable after $\sim 2.5$ Gyr. Hence there is some flexibility in the choice of
the disk surface density that can lead to bar formation, even 
for a highly concentrated halo, but always provided that the
disk is sufficiently massive.

\subsection{Gasdynamical runs}

The simulations employing an exponential gaseous disks with $f_d=0.05$
and an isothermal equation of state show results that are sensibly 
different from those of the corresponding collisionless runs. 
A distinct bar forms
within a $c=4$ halo (Lmd1c4g, see Figure 7) after 5 Gyr, 
while only a short-lived bar-like distortion
was observed in the corresponding collisionless run and only for the 
lowest initial value of $Q$ (model Lmd1c4Q3, see Table 1). 
Only milder non-axisymmetric patterns, either
spiral-like or oval, appear after a few Gyr in halos having $c=12$ (Lmd1c12g). 
Later these patterns evolve into
clump-like structures near the peaks of the surface density distortions. 
We note that the bar which forms in the Lmd1c4g is visibly
shorter than that produced in any of the stellar-dynamical models (its
radius is $\sim 0.3 R_h$) and it
also appears much later; we intend to explore in detail the formation
path of this type of gaseous bar in a forthcoming paper. Once formed, the bar quickly evolves (in 1-2 Gyr, of order of its rotation period)  
into something resembling an oval distortion and then a 
bulge-like component (Figure 8). It is thus a really short-lived feature during the evolution of the gaseous disk. Rising the initial disk temperature by
about $50 \%$ (model Lmd1c4gb in Table 1) is enough to prevent the growth
of the bar.
Nonetheless, this simulation suggests that
cold gaseous disks appear more prone to undergo non-axisymmetric instabilities 
with respect to their stellar analog (see also Figure 5 on the ellipticity).
The reason is that the temperature is
constant with time for model Lmd1c4g while its stellar dynamical
analog, the velocity dispersion, is
not constant in the corresponding collisionless models (e.g. Lmd1c4), 
instead it actually increases with time
, driving $Q$ towards large values. For example, after 4 Gyr (just before 
the bar appears in model Lmd1c4g) the average 
velocity dispersion of model Lmd1c4 has increased by a 
factor 2.5 within the disk scale length in response to weak spiral 
instabilities, and $Q > 2$ results throughout the disk, too high for the
bar to develop. We tested whether the gaseous disk would respond in a similar
way if cooling was inefficient; we thus evolved the same disk adiabatically
for 6 Gyr and we found that, as expected, the bar does not form in this case.
In fact, after 4 Gyr, the temperature has grown by more than a factor of
5 in the inner few kpcs (reaching 35000 K), which raises $Q$ by a about
a factor of 2 ($Q \propto T^{1/2}$). The disk now looks even smoother
than the stellar disk, i.e. even less non-axisymmetric structure is present 
(compare Figure 7 and Figure 3).
However, the isothermal calculations are probably more realistic 
at least in a global sense as 
cooling should be efficient enough to keep the temperature of the HI 
disk below  $10.000$ K (Martin \& Kennicutt 2001 - see also section 2). Therefore, we tend to 
conclude that in reality gaseous disks should be more susceptible to grow
bars compared to stellar disks, although this does not imply anything on
the relative longevity of the two types of bars (see below). 
The disk evolution is
also dependent on the type of the gas profile.
In models where the gaseous disk has a constant gas-density profile (and is
evolved isothermally) a strong bar does not appear neither in low nor in 
highly 
concentrated halos. In these models the surface density of the gaseous disk
is too low everywhere (see the rotation curves in Figure 2 and the values
of the stability parameters in Table 1) and therefore 
it is not surprising that the growth of bar modes is suppressed.

\begin{center}
\begin{figure}
\epsfxsize=8truecm
\epsfbox{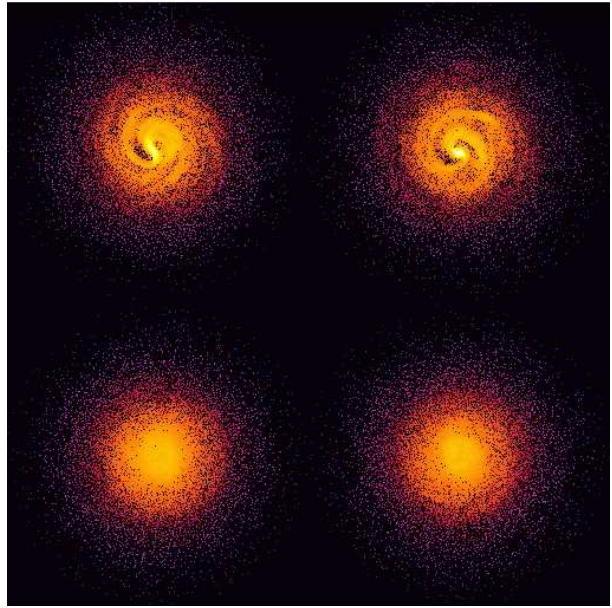}
\caption{Color-coded face-on view of the stellar density of model 
Lmd2c4g. Brighter
colors represent higher densities. Snapshots of both the isothermal
run (top) and the adiabatic run (bottom) are shown at 5 Gyr (left)
and 8 Gyr (right). Boxes are 30 kpc on a side.}
\end{figure}
\end{center}

\begin{center}
\begin{figure}
\epsfxsize=8truecm
\epsfbox{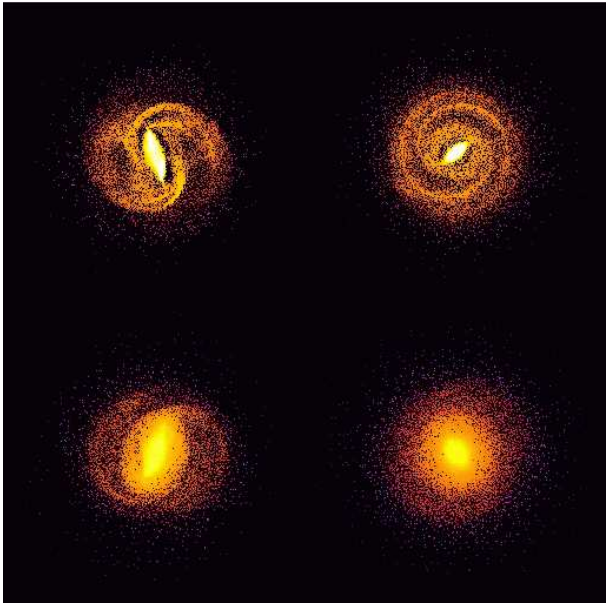}
\caption{Color-coded face-on view of the stellar density of model 
Lmd2c12g. Brighter
colors represent higher densities. Snapshots of both the isothermal
run (top) and the adiabatic run (bottom) are shown at 3 Gyr (left)
and 6 Gyr (right). Boxes are 30 kpc on a side.}
\end{figure}
\end{center}

Massive gaseous disks ($f_d=0.1$, models Lmd2c12g or Lmd2c4g) become strongly bar unstable in a similar fashion as their stellar counterparts (Figure 8). 
The bar instability appears even when the same models are evolved using an 
adiabatic equation of state. The instability in massive gaseous disks is
"dynamical" as in the collisionless case - it arises early, on a timescale 
comparable to the dynamical time of the disk (Figure 8). 
The bar morphology evolves faster than in the stellar dynamical case, 
resembling more an oval distortion after just one rotation period 
($\sim 3$ Gyr, see Figure 8). This 
difficulty shown by the gaseous disks in sustaining a barred potential was 
also reported by Friedli \& Benz (1993) who found that the periodic 
$x_1$ orbits that
support a stellar bar are quickly destabilized by shocks developing 
in their loops. Shocks are indeed clearly visible along the bar edges in
our gasdynamical runs and the elongation of the density distribution 
decreases rapidly, especially in the adiabatic runs, where a higher pressure  
develops that smears out the density perturbation (see Figure 5).

Christodolou et al. (1995) have extended the stability criterion based on $\varepsilon_d$ to gaseous disks within halos, deriving $\varepsilon_d \simgt 0.9$
as a condition for stability to bar formation in the latter systems. 
Their result
was not based on numerical simulations of disk+halo systems but was obtained
from a comparison between quite idealized ellipsoidal fluid configurations and  their stellar analogs.
Our numerical results confirm that gaseous models are slightly more stable 
than stellar disks if we compare models starting with a similar initial 
$Q$ and $\varepsilon_d$ and there is no cooling (as in the adiabatic case),  
However, in general $\varepsilon_d \simgt 1$ for stability, i.e. we obtain a 
criterion slightly stronger than for stellar disks if we take into account 
the effect of cooling (isothermal case). Nonetheless, the fact that gaseous
bars appear to weaken sooner than stellar bars (this being also dependent
on the equation of state) suggests that 
their overall dynamical impact on the galaxy (including the halo) might be
less important.

\begin{center}
\begin{figure}
\epsfxsize=8truecm
\epsfbox{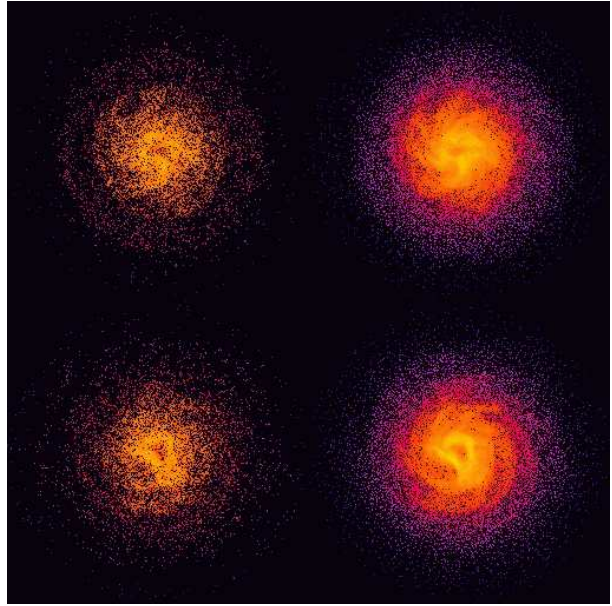}
\caption{Color-coded face-on view of the stellar density  of model 
Lmd1c4sg (see Table 1). Brighter colors represent higher densities. Snapshots are taken after
3 Gyr (top) and 7 Gyr (bottom). The stellar distribution is shown on the left, the gas on the
right. Boxes are 30 kpc on a side.}
\end{figure}
\end{center}

In the model where stars and gas contribute equally to the disk (model Lmd1c4gs)each of the two baryonic components  
has a surface density reduced by 50\% compared to the corresponding models
having a single baryonic component (Lmd1c4 or Lmd1c4g) and a bar does not form 
regardless of the halo concentration (Figure 9).
Analytical
calculations and numerical experiments have shown that a disk made of stars
and gas can be more susceptible to gravitational instability than either of its individual components would (Jog 1992; Elmegreen 1995; Jog 1996);
however, those results apply when
at least one of the components is marginally unstable (yielding $Q \sim 1$),
whereas in  the present case the surface density is very low for both
components and, as a consequence, all the relevant parameters have values
well within the regime of stability (see Table 1).

\subsection{Morphological evolution of the bars}

In models that go bar unstable (both stellar and gaseous) we always observe a
morphological evolution from a bar into a bulge-like central structure over a few bar dynamical times (a few Gyrs, see Figure 10 and 11),
in agreement with the secular evolution scenario (Combes \& Sanders 1981; Combes et al. 1990; Merritt \& Sellwood 1994; Carollo et al. 1999).
In the latter vertical instabilities occur locally as a consequence of 
resonances between the bar rotation speed and the vertical oscillation frequency 
of the stars (Combes \& Sanders 1981) or simply result because the orbital families
supporting a flat, radially anisotropic bar become unstable to bending modes (Pfenniger 1984; 1985; Raha et al. 1991; Sellwood \& Merritt 1994). 
When viewed edge-on the galaxies go from a disky shape to 
peanut-shaped soon after the bar forms and become progressively rounder
(Figure 11). As the rotating bar evolves stars also lose angular momentum to the 
surroundings and lead to an increased surface density. The increase in central
density is more pronounced in the gaseous disk (compare Figure 12 and 
Figure 13). In fact, in addition to the  gravitational losses already present in 
collisionless systems, some dissipation of angular momentum can 
occur in the shocks near the bar edges; in the adiabatic runs this
is compensated by the subsequent heating and expansion of the gas and in
the end the inner profile looks more similar to that of the collisionless runs.
The bar persists
for several Gyr, although it weakens as time goes on (the elongation of the
density distribution diminishes, see last snapshot of Figure 10), maybe
because of the growing central mass concentration (Hasan \& Norman 1990; 
Friedli \& Benz 1993; Norman, Sellwood \& Hasan 1996).
An early-type spiral results, but one in which the disk has still a very low 
surface density.
The central bulge has an exponential profile with a scale length 5-10 times
smaller than that of the surrounding disk (Figures 12, 13 and 14). 
The disk is still exponential
but considerably flatter than at the beginning. Its scale length more than doubles,
reaching as much as 9 kpc in one case; the surface density decreases correspondingly and would lead to 
a disk central surface brightness a few magnitudes
lower than in the initial conditions (Figure 12, 13).
These and other properties of the final states of our bar-unstable models closely
match those of the red early-type LSB galaxies studied by Bejersbergen et al. (1999). 
In the next section we will discuss in detail how far the comparison with such 
class of galaxies can be pushed.

\begin{center}
\begin{figure}
\epsfxsize=8truecm
\epsfbox{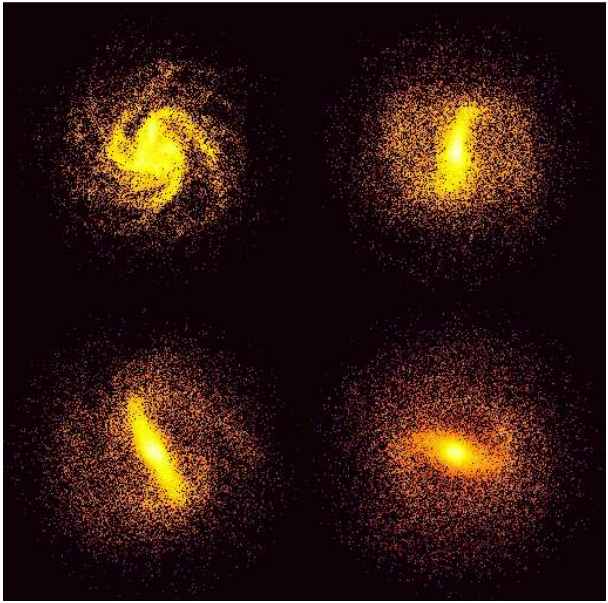}
\caption{Color-coded face-on view of the stellar density of model Lmd2c12. Brighter
colors represent higher densities. From the upper left
to the bottom right, snapshots are shown at, respectively, 1,3,4.5,7 Gyr are 
shown.Boxes are 25 kpc on a side}
\end{figure}
\end{center}

\begin{center}
\begin{figure}
\epsfxsize=8truecm
\epsfbox{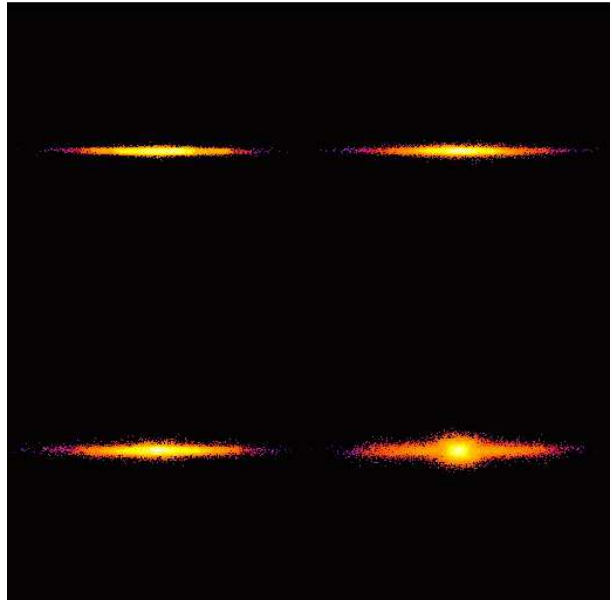}
\caption{Color-coded edge-on view of the stellar density of model Lmd2c12. Brighter
colors represent higher densities. From the upper left
to the bottom right, snapshots are shown at, respectively, 1,3,4.5,7 Gyr are 
shown. Boxes are 25 kpc on a side}
\end{figure}
\end{center}

Our simulations do not include star formation.
We might ask 
how long will a galaxy remain in a mostly gaseous state as assumed for some
of our models.
Kennicutt (1998) and Martin \& Kennicutt (2002) have shown that the 
Toomre Q parameter gives a good indication of the local density threshold 
above which star formation can occur. Observations of galaxies suggest
that  $Q  < 1.44$ is required for star formation to proceed. Our gaseous 
disks satisfy the latter criterion nearly everywhere (except in the inner few
hundred parsecs) but the star formation
rates calculated from the local initial gas surface density as in Kennicutt (1998) 
would be  as low as $ \simlt 0.3 M_{\odot}/yr$ within 10 kpc (corresponding
to a few disk scale lengths) for models with $f_d=0.05$.
Such modest star formation rates fall near the lowest measured
by Kennicutt (1998) for spiral galaxies and are comparable to the
star formation rates that Gerritsen \& de Blok (1999) obtain in their
simulations of LSB galaxies. Therefore, such a galaxy will remain mostly
gaseous for a good fraction of its lifetime. In addition, these low star 
formation rates imply a weak supernovae feedback with little impact on the 
thermal and  dynamical evolution of the gas, this being consistent with 
our assumption of an isothermal evolution.
Of course, when significant non-axisymmetric structure, bars or clump-like
features form all these assumptions could break locally as  the
surface density grows above some threshold.
In the light disk models ($f_d=0.05$) the surface density
increases, on average, by only a factor 2 in the bar region (in the inner
few kpcs in Figure 14). 
Instead, in models with heavy disks the average surface density within
1-2 scale lengths increases by a factor of 10 and would result in a typical 
star formation rate of  $> 20 M_{\odot}/$yr, comparable to that of actively 
star forming HSB galaxies (Bell et al. 1999).
The highest star formation rates would occur in the center, inside
the growing bulge-like component. As we mentioned,
we believe that the final state of the heavy disks resembles 
the red LSB galaxies studied by Bejersbergen et al. (1999).
Clearly more quantitative predictions of the luminosity and color evolution
of the simulated galaxies demand that we include star formation and the mechanisms
that can regulate it, for instance heating by the UV radiation from hot stars and supernovae
(Gerritsen \& Icke 1997;Gerritsen \& de Blok 1999).
Simulations including these additional mechanisms will be the subject of
a forthcoming paper.
Nonetheless, our main focus here is on the initial development of 
non-axisymmetric structure
in both stellar and gaseous disks and in this respect LSB models start 
with surface
densities low enough to justify neglecting star formation and 
its effects.

\section{Discussion}

We have shown that bar formation in low surface brightness stellar disks is 
unlikely
unless they are more massive than usually believed. Light disks generate
only transient non-axisymmetric distortions even when starting from a
state with $Q < 1$. 
Instead, purely gaseous disks can become bar unstable even for quite 
low masses  provided that their halos have
low concentrations and their temperature stays low enough to maintain
$Q \simgt 1$
(as shown in the simulations adopting an isothermal equation of state).
Growing modes are damped more easily in stellar disks because the 
velocity dispersion increases in response, rising $Q$ to values well 
above 1; gaseous disks behave similarly only when the cooling is completely 
switched off (like in the experiments with an adiabatic equation of state). 
Although adopting an isothermal equation of state might seem simplistic, the temperatures we need to keep $Q \sim 1.2$ over most of the radial extent of our 
models, $T = 7500 K$, is typical of the HI component of spiral galaxies 
(Martin \& Kennicutt 2001).

\begin{center}
\begin{figure}
\epsfxsize=8truecm
\epsfbox{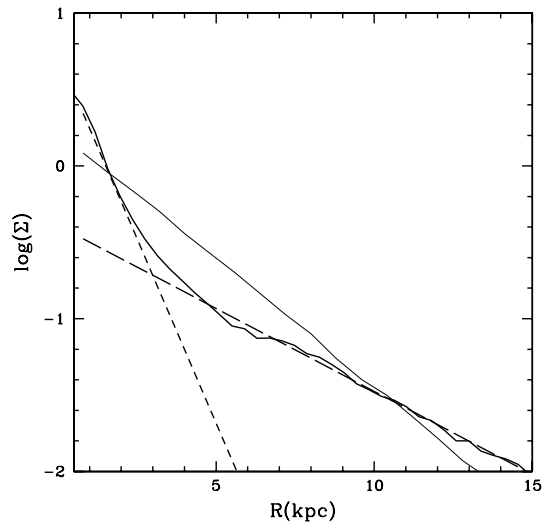}
\caption{The final stellar surface density profile (thick solid line) of model Lmd2c12 (see Table 1) is plotted against 
the intial purely exponential profile (thin solid line). Exponential  fits to the inner (short-dashed 
line) and outer 
(long-dashed line) part of the profile are shown, with scale lengths of, respectively, 900 pc and
4 kpc (while the initial disk scale length was $\sim 3$ kpc).
The surface density is expressed in the units of the
simulation}. 
\end{figure}
\end{center}

Concentrations for halos forming at a redshift $\ge 2$  in LCDM models 
are sufficiently low ($c < 6$ at $V_c < 80$ km/s) to include
even the values required to drive bars in light gaseous disks
(Bullock et al. 2001).
The combination of a cold, mostly gaseous disk and a low concentration 
halo was probably not uncommon at high redshift and perhaps so were barred 
LSB galaxies. 
However, when we look at the present-day distribution of halo concentrations 
in cosmological simulations
(this is basically the combination of the distributions for objects
formed at {\it any} redshift) we see that such low values are found for only
a few percent of the systems (Eke, Navarro \& Steinmetz 2001; Bullock et al. 2001). 

Therefore an even stronger result of our work is that
halos as concentrated as typically occurs in LCDM models (in passing we note 
that even the highest value of the halo concentration considered here, $c=12$, is still conservative) inhibit bar formation,
even when a satellite as massive as the Large Magellanic Cloud is perturbing the disk during several close passages.

On the other end, massive disks only 10\% lighter than the dark matter halo
become bar unstable regardless of halo concentration but the resulting bar is longer (relative to the initial disk scale-length) and, possibly,
stronger for higher halo concentrations (see section 3.1).
This is consistent with the findings
of Athanassoula (2002) and Athanassoula \& Mistriotis (2002), who
claim that a higher halo mass within the disk region produces
longer and stronger bars because these lose more efficiently angular momentum 
through resonant exchange with halo particles (but see also Athanassoula 2003).

\begin{center}
\begin{figure}
\epsfxsize=8truecm
\epsfbox{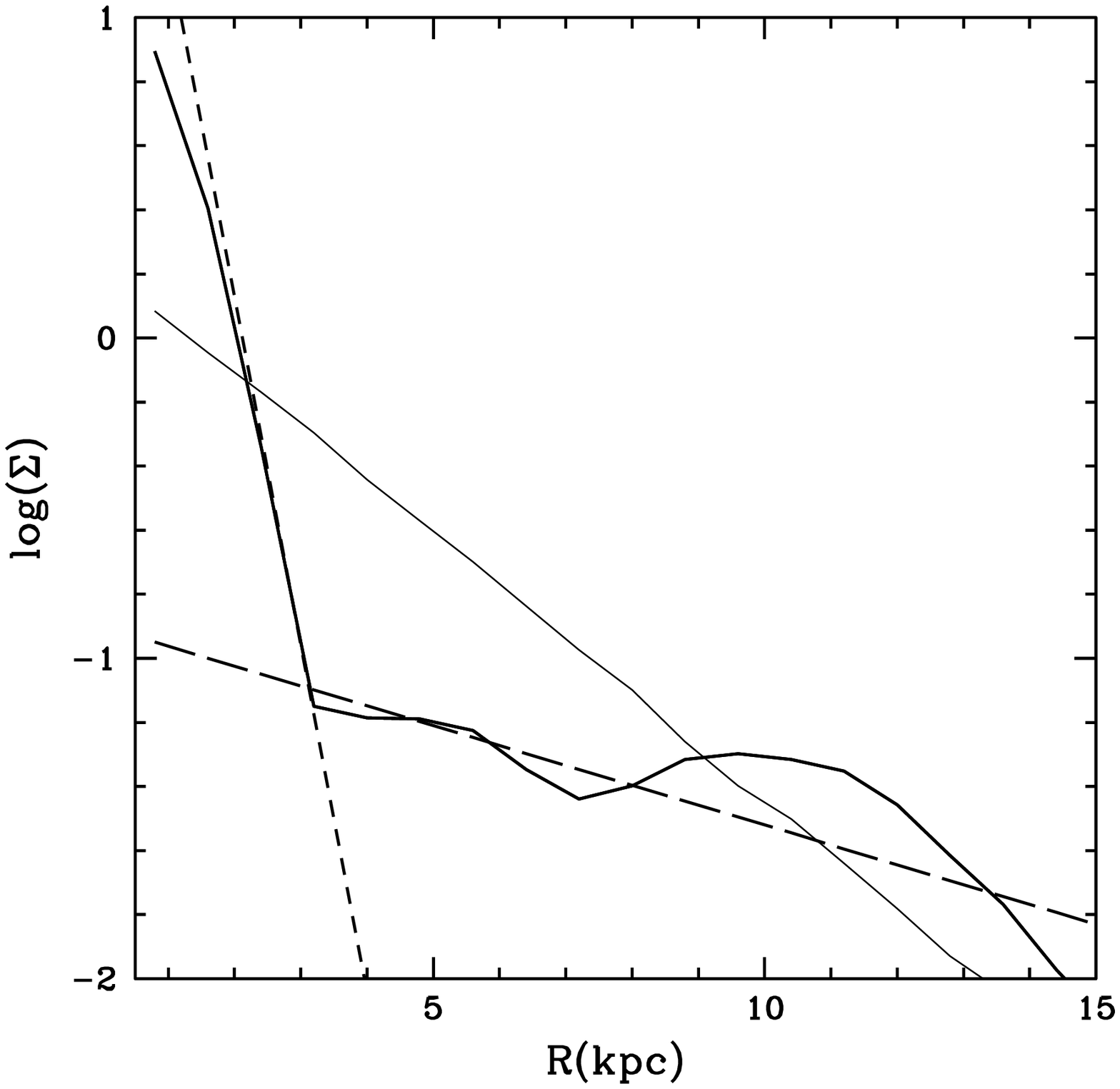}
\caption{ The final gas surface density profile (thick solid line)
of model Lmd2c12g (see Table 1) is plotted against the initial
purely exponential profile (thin solid line). The results for the isothermal run
are shown.
Exponential  fits to the inner (short-dashed line) and outer 
(long-dashed line) part of the profile are shown, with scale lengths of, respectively, 800 pc and
9 kpc (while the initial disk scale length was $\sim 3 kpc$).
The surface density is expressed in the units of the simulation}
\end{figure}
\end{center}

As explained in section 3.3, bars evolve into bulge-like structures in
a few Gyr in both stellar and gasdynamical runs. 
The bulge reaches a surface density
significantly higher than the rest of the disk (Figures 12, 13, and 14); 
in models with gaseous bulges 
a mean star formation rate $\sim 1 M_{\odot}/yr$ would result in light disk models following Kennicutt (1998) - this implies fairly rapid gas consumption timescales, 
$\simlt 1$ Gyr.
In contrast, the rest of the disk will continue to form stars with a rate
from 10 to 20 times lower and will eventually keep a sufficient gas 
reservoir to form stars until the present epoch, thus maintaining 
rather blue colors as those of many observed LSBs. 
Star formation rates between 10 and 30 times higher would result in the bulges
forming from
massive disks; in this case a real starburst will likely convert most of 
the gas into stars in $\sim 10^{8} years$.
In the latter case the bulge B band surface brightness would be 
comparable to that of bulges of normal early type spirals, $\mu_B \sim 19$ mag arcsec$^2$, for a stellar-mass-to-light ratio $\sim 1$ (typical of a young
stellar population) based on the stellar or gas surface density measured 
in our simulations. The bulge would eventually fade following the burst;
neglecting any further evolution of the surface density, if the stellar
population undergoes passive evolution and reaches the typical 
stellar mass-to-light ratios of spheroids after a few Gyr, ${(M/L_B)}_*
\sim 3-4$, the final
central B band surface brightness will be $\simgt 20$ mag arcsec$^{-2}$
((these numbers are nearly independent
on whether we consider the case of the models with only a stellar component 
or we compute surface densities from the models with gas assuming that this
has been turned into stars).
Such systems would resemble the red LSBs
studied by Bejersbergen, de Blok \& Van der Hulst (1999). The latter have a low surface brightness disk with central
red bulges similar in structure to those of HSB galaxies but slightly fainter than them. Their bulge-to-disk ratios ($B/D$) in the I band (which should provide
a good measure of the actual mass ratios) vary between less than
0.05 to as much as 0.5 and their total (disk + bulge) $B$ band luminosity 
is $17 < M_B <21$. The final states of our models would have 
comparable luminosities, $18.5 < M_B < 19.5$ (for 
${(M/L_B)}_*=2$) and the bulge contributes from $15$ to $50 \%$ of the total
mass of the baryons (we measure the mass of the bulge as the mass contained
within the radius at which the slope of the stellar profile steepens
significantly).
A similar evolutionary scenario has been recently proposed by Noguchi (2002)
to explain giant LSB galaxies which bear a resemblance to the red LSBs
but have a much larger size, bigger $B/D$ ratios and even redder colors 
(Sprayberry et al. 1995).
However, while Noguchi starts from a model of a normal HSB galaxy 
(without gas) 
here we have shown that even galaxies that start with fairly low
surface density disks can become bar unstable and form a bulge provided 
that their disk is sufficiently massive with respect to the halo.
The bulges that form in our models have both
profiles and scale lengths that match those of the
red, early-type LSB galaxies in  Beijersbergen et al. (1999).
; they are fit by exponential laws
and their scale lengths are typically 
$\sim 0.1-0.2 R_h$ (Figure 13, 14, and 15). 
When translated into physical sizes,
bulge scale lengths are as large as 0.4-0.9 kpc, i.e significantly larger 
than the bulges of "normal" HSB galaxies of similar luminosity (de Jong
\& Van der Kruit 1994). Being the product of secular disk evolution, the bulge scale 
lengths are correlated with those of the surrounding LSB disks, matching another
feature of the galaxies in  Beijersbergen et al. (1999)
(for HSB galaxies a similar correlation
is indeed found, see de Jong \& Van der Kruit 1995 and MacArthur, Courteau \& Holtzman 2003).
Also, as already mentioned in section 3.3, disks remain exponential but increase their scale length by at least
a factor of 2, reaching as much as 9 kpc (e.g. Figure 13).
Red early-type LSBs also
have disks with huge scale lengths, bigger than their blue counterparts for the
same luminosity.
Some of the red LSBs also exhibit significant spiral structure; this appears to 
be triggered
by the bar in our simulations and persists even after this has turned into a bulge,
especially for purely gaseous disks (Figure 7 and 10).

In our scenario bulges form after a (gaseous) disk has already 
settled into the halo as a result
of  secular bar evolution. The bulge would be a younger than the
disk in a dynamical sense but after a few Gyr it would look redder than 
the latter
due to the different mean age of its stars; indeed, as we showed above, 
the different density
structure of the two components suggests that, while an early burst of 
star formation is 
plausible for the bulge, the disk would undergo a weaker but more 
prolonged star formation.
In particular, the spiral arms might excited by the central bar would play
an important role in the star formation history of the disk.
Bars and rings are actually present in some of the red LSBs, in which case 
they also have bluer colors in their central part; these systems resemble
the intermediate stages seen in our simulations, while
systems with distinct bulges are likely in the late evolutionary stage, and
indeed they also look redder in the center as if the stellar population there
has already undergone significant fading.
Future detailed
observations of color and age gradients throughout the disk and the bulge should enable to better judge how realistic is the scenario proposed here.
Simulations including star formation and explicit
cooling and heating are better suited to explore further this issue and will be
the subject of a forthcoming paper. 

How large a population of LSB galaxies with bars/bulges should we expect 
based on our model? 
The naive expectation is that
such galaxies should be rare as they originate from systems with disk mass
fractions close to the upper limit for the baryon fraction (Jaffe et al. 2001).
On the other end, several authors, by modeling galaxy rotation curves, find
a positive correlation between the spin parameter and the disk mass fraction 
(e.g. Burkert (2000), Van den Bosch et al. (2001), Jimenez, Verde \& Oh (2003)),which would imply that massive disks are more common among LSB galaxies.
This correlation is not well understood. Burkert (2000) has proposed that it
might result from a correlation between disk specific angular momentum (which
might be different from that of the parent dark matter halo) and the disk mass 
fraction - however at the moment it cannot be excluded that the correlation is 
produced by systematic errors in the procedure used to determine the
dark halo parameters, especially when the rotation curves do not
extend far enough in radius (Verde et al. 2002). Nonetheless, 
the recent analysys of 400 rotation curves reported in Jimenez et al. (2003)
indicates that more than $50\%$ of high spin systems (i.e. with $\lambda 
\ge 0.065$, the minimum value considered in this paper) have disk mass 
fractions near $0.1$ (L. Verde, private communication). These results leave
open the interesting possibility that many LSB galaxies are massive enough
to become bar unstable and undergo the morphological transformation described
in this paper.

\begin{center}
\begin{figure}
\epsfxsize=8truecm
\epsfbox{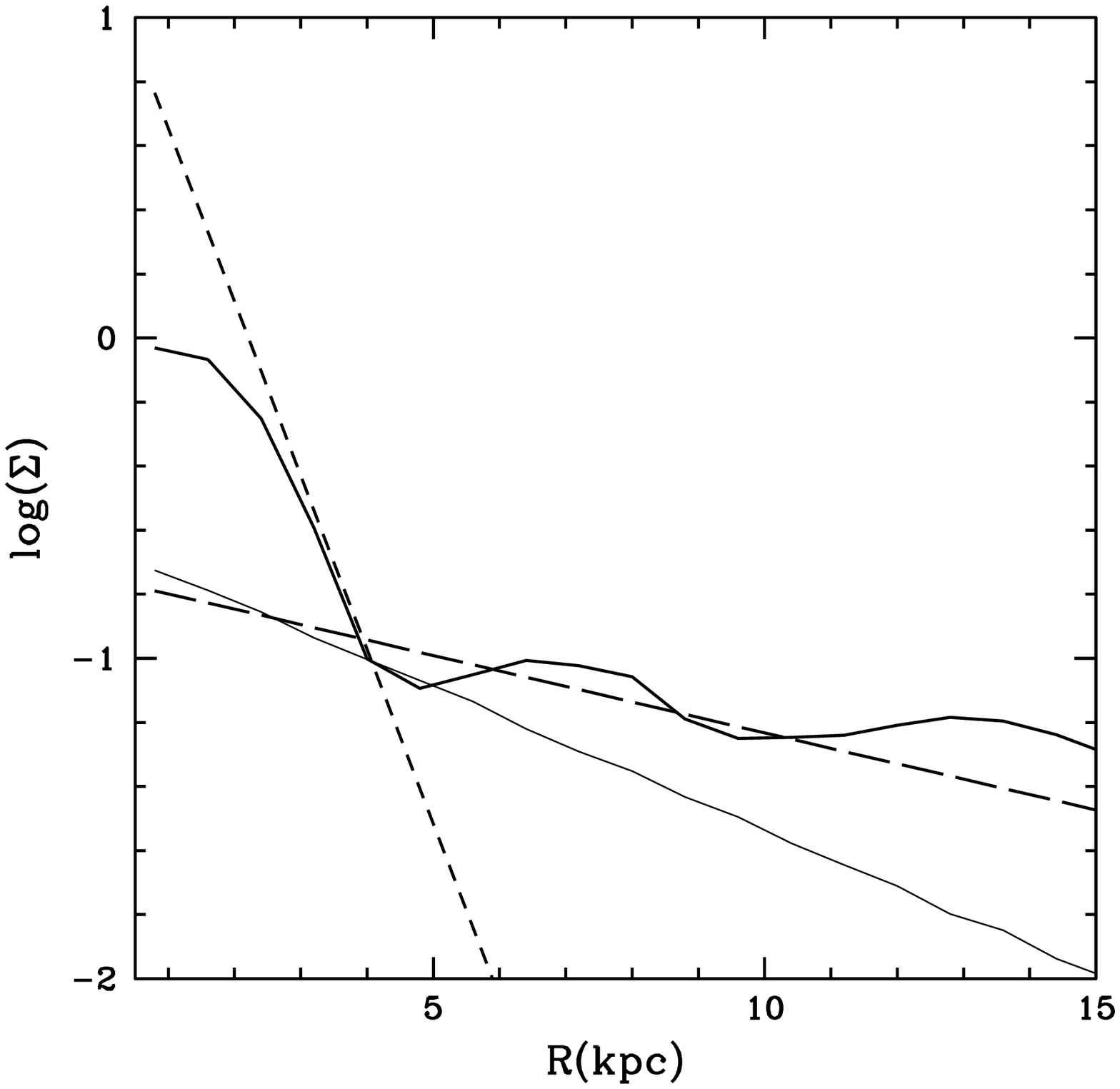}
\caption{The final gas surface density profile (thick solid line) 
of model Lmd1c4g (see Table 1) is plotted against 
the intial
purely exponential profile (thin solid line). The results for the isothermal
run are shown.
Exponential  fits to the inner (short-dashed line) and outer 
(long-dashed line) part of the profile are shown, with scale lengths of, respectively, 400 pc and 7
kpc (while the initial disk scale length was about 5 kpc).
The surface density is expressed in the units of the
simulation}. 
\end{figure}
\end{center}

Relating the final state of bar unstable but light ($f_d < 0.1$) gaseous 
disks to  known galaxies turns out to be less obvious. 
The bulge-like component that appears late in these systems would have an
unusually 
low optical surface brightness as it forms out of gas having very low 
surface densities; assuming ${(M/L_B)}_* = 4$, as typical of 
old spheroid stellar populations, these bulges would have a B band surface 
brightness 
$\sim 24$ mag arcsec$^{-2}$. 
Such a low surface
brightness is significantly lower than any of those of the objects studied 
in Beijersbergen et al. (1999). However, although the bulge will be fainter 
than the disk and might be hardly recognizable in optical bands, 
it should still stand out when observed with longer wavelengths because
of its intrinsically higher stellar density (see Figure 14).
Many blue LSB galaxies or LSB dwarfs do not have significant 
spiral/non-axisymmetric structure, appearing rather amorphous or
irregular (de Blok,
Van der Hulst \& Bothun 1995); on the
basis of our results, the simplest interpretation of these systems is that 
they are stable because they have fairly light disks, $f_d <0.1$ and/or
halos as highly concentrated as expected in a LCDM model.   
Most of the galaxies whose measured rotation curves are nourishing
the debate on dark matter cores would fall into the latter category; 
therefore one would be tempted to conclude
that bars cannot occur in these systems and hence cannot be invoked
to explain the present-day structure of their inner halo. 
However, we cannot exclude
that the available observations lack sufficient resolution and are still 
hiding
some important clues to the past dynamical histories of these objects
(see de Blok, Walter \& Bell 1999).
New high-resolution observations of individual blue LSB galaxies at long 
wavelengths will show whether dim, red bulges are present in galaxies 
previously classified as blue, late-type 
LSBs and hence whether bar formation could have taken place in
the past. Only if such spheroidal components will turn out to be 
extremely rare we will conclude that  bar formation, and thus bar/halo 
interactions, can be neglected for these galaxies. 
Interestingly, the recent observations in J and K bands
by Galaz et al. (2002) suggest that at least a fraction of the systems that 
appear featureless in the optical hide a central red bulge or 
nucleus when observed in the near-IR. 
Of course, high-resolution images might
also reveal bar-like distortions.
Indeed, some bars are observed  in a few LSB galaxies included in
a recent sample used by
Swaters et al. (2003) to measure H$\alpha$ rotation curves; these authors claim
that the barred systems
have slightly shallower halo inner slopes compared to the non-barred
ones, although they admit that non-circular motions due to the bars themselves
make the determination of the actual rotational velocity very hard
and questionable in these cases.

However, even though evidence of present or past 
bars will be found in a number of LSB galaxies, our simulations
do not provide strong support for a scenario in which bars can
significantly affect the structure of the dark halo.
In fact, the bars arising in our simulated galaxies are rather short, 
from 3 to 8 times smaller than the halo scale radius, $r_s$ (see
section 2), hence typically 
shorter than the long massive bars assumed in Weinberg \& Katz (2002) (but note that the bar in model Lmd2c12b, which is the longest due to the
large value of the spin parameter, approaches the size of the bar considered
in that paper);
recent calculations by Sellwood (2002) show that, no matter how effective
is the transfer of angular momentum between the bar and the halo, bars 
up to 4 times smaller than $r_s$ do not 
store enough angular momentum to produce the observed large dark matter cores. 
A particularly short bar arises in light disks (model Lmd1c4g), its size 
being $> 10$ times smaller than $r_s$, hence smaller 
than any of the bars studied by Sellwood (2002). Finally, the fact that
all the bars in our simulations, and especially the
gaseous ones, seem to undergo a fast morphological evolution, also
goes in the direction of a modest dynamical impact on the dark halo.

\section{Summary}

In this paper we have shown that bar formation in LSB galaxies is possible, yet the
initial conditions required for this to happen apply to only a narrow region of the
parameter space made available by currently favored
 hierarchical models of structure formation. In addition,
the characteristics of the bars that eventually form in such systems as well 
as their rapid morphological evolution
hardly support the idea that bar-halo interactions play
a crucial role in shaping the inner structure of the dark matter halo.
Our main findings are summarized as follows;

\begin{itemize}

\item

The halo/disk mass ratio within the region where the disk lies determines
whether the studied LSB models are stable or not. This ratio is fixed
once both the halo concentration and the disk mass are fixed and sets
the degree of swing amplification of $m=2$ modes. In particular, LSB galaxies 
with disks as massive as 10\% of the halo mass can become bar unstable 
for quite a range of concentration values.

\item

LSB galaxies with (typical) light disks ($f_d < 0.1$) can become bar unstable 
if their halo concentration is as low as $c=4$ and their disks are essentially 
gaseous and cold. Such low concentrations are rare among LCDM halos.

\item

Bars forming in LSB galaxies are usually much shorter than the halo scale
radius - they are not expected to contain enough angular momentum to 
turn the inner cusp of dark matter halos into a core.

\item

Both gaseous and stellar bars evolve into central bulge-like structures after a few Gyr.
The bar and the bulge have surface densities up to 2 order of magnitudes higher
 than the surrounding low surface density disks. Radically different star
formation histories are thus expected in the central part of the galaxy as
opposed to the extended disk.

\item

The structural properties of the final states (after up to 10 Gyr of evolution) of bar unstable, massive
LSB disks are consistent with those of the red, early-type LSB galaxies observed by
Beijersbergen et al. (1999). Our secular evolution scenario naturally
explains observed correlations like that between the disk and bulge scale
lengths.

\item

Blue LSB galaxies included in samples used to measure rotation curves
usually appear featureless in the optical; our findings
imply that that a dim, red bulge-like component must be present at
their center if they ever formed a bar. 
Future observations of these galaxies in the near infrared
should reveal whether these hidden bulges exist in most LSBs;
first attempts in this direction (Galaz et al. 2002) suggest that
this might be the case. 

\end{itemize}

\section{Acknowledgements}

L.M. is grateful to Victor Debattista for carefully reading an early version 
of the paper and for providing plenty of useful comments and suggestions. We
also thank James Schombert, Martin Weinberg, Neal Katz and Thomas Quinn for 
interesting and stimulating discussions on LSB galaxies and on the dynamics 
of barred galaxies, and Marcella Carollo, Julianne Dalcanton and Licia Verde 
for helpful information on recent observations. The simulations were 
performed on the 64 processor Intel cluster at the University of Washington, 
on the Compaq-Alpha cluster at the Pittsburgh Supercomputing Center and on the 
ORIGIN 3800 at CINECA.

\end{document}